\documentclass[11pt,twoside]{article}
\usepackage[makeroom]{cancel}
\usepackage{latexsym}
\usepackage{longtable}
\usepackage{epsfig}
\usepackage{graphicx,bbm,psfrag}
\usepackage{amssymb,amsmath,amsthm,booktabs,mathtools}
\usepackage{bm}
\usepackage{color}

\newtheorem{theorem}{Theorem}[section]

\newtheorem{lemma}[theorem]{Lemma}
\newtheorem{corol}[theorem]{Corollary}

\newcommand{\bc}{\begin{center}}
\newcommand{\ec}{\end{center}}
\def\ba#1{\begin{array}{#1}\displaystyle}
\newcommand{\ea}{\end{array}}

\newcommand{\beq}{\begin{equation}}
\newcommand{\eeq}{\end{equation}}
\newcommand{\beqa}{\begin{eqnarray}}
\newcommand{\eeqa}{\end{eqnarray}}
\newcommand{\no}{\nonumber}
\newcommand{\n}{\nonumber\\}
\newcommand{\bi}{\begin{itemize}}
\newcommand{\ei}{\end{itemize}}

\def\lt#1{\left#1}
\def\rt#1{\right#1}
\def\t#1{\tilde{#1}}

\def\frc#1#2{\frac{#1}{#2}}

\newcommand{\p}{\partial}

\newcommand{\vac}{{\rm vac}}
\newcommand{\bra}{\langle}
\newcommand{\ket}{\rangle}
\newcommand{\Z}{{\mathbb{Z}}}
\newcommand{\N}{{\mathbb{N}}}
\newcommand{\R}{{\mathbb{R}}}

\newcommand{\Tr}{{\rm Tr}}

\newcommand{\Or}{{\cal O}}

\newcommand{\ep}{\epsilon}

\newcommand{\ri}{{\rm i}}
\newcommand{\re}{{\rm e}}
\newcommand{\rl}{{\rm l}}
\newcommand{\rr}{{\rm r}}
\newcommand{\dd}{{\rm d}}

\newcommand{\dbra}{\langle\hspace{-0.17em}\langle}
\newcommand{\dket}{\rangle\hspace{-0.17em}\rangle}

\newcommand{\halmos}{\rule{1ex}{1.4ex}}
\newcommand{\eproof}{\hspace*{\fill}\mbox{$\halmos$}}

\begin{document}

\begin{titlepage}

\begin{center}
{\Large {\bf Entanglement Content of Quantum Particle Excitations III.  \\[0.2cm]  Graph Partition Functions}}

\vspace{0.8cm} 
{\large \text{Olalla A. Castro-Alvaredo${}^{\heartsuit}$, Cecilia De Fazio${}^{\clubsuit}$, Benjamin Doyon${}^{\diamondsuit}$ and Istv\'an M. Sz\'ecs\'enyi${}^{\spadesuit}$}}

\vspace{0.8cm}
{\small
 ${}^{\heartsuit\, {\clubsuit}  \, \spadesuit}$ Department of Mathematics, City, University of London, 10 Northampton Square EC1V 0HB, UK\\
\vspace{0.2cm}
{${}^{\diamondsuit}$}Department of Mathematics, King's College London, Strand WC2R 2LS, UK}\\

\end{center}

\vspace{1cm}
We consider two measures of entanglement, the logarithmic negativity and the entanglement entropy, between regions of space in excited states of many-body systems formed by a finite number of particle excitations. In parts I and II of the current series of papers, it has been shown in one-dimensional free-particle models that, in the limit of large system's and regions' sizes, the contribution from the particles is given by the entanglement of natural qubit states, representing the uniform distribution of particles in space. We show that the replica logarithmic negativity and R\'enyi entanglement entropy of such qubit states are equal to the partition functions of certain graphs, that encode the connectivity of the manifold induced by permutation twist fields. Using this new connection to graph theory, we provide a general proof, in the massive free boson model, that the qubit result holds in any dimensionality, and for any regions' shapes and connectivity. The proof is based on  clustering and the permutation-twist exchange relations, and is potentially generalisable to other situations, such as lattice models, particle and hole excitations above generalised Gibbs ensembles, and interacting integrable models.
\medskip

\vfill

\noindent 
${}^{\heartsuit}$ o.castro-alvaredo@city.ac.uk\\
{${}^\clubsuit$} cecilia.de-fazio.2@city.ac.uk\\
{${}^{\diamondsuit}$} benjamin.doyon@kcl.ac.uk\\
{${}^{\spadesuit}$} istvan.szecsenyi@city.ac.uk\\

\hfill \today

\end{titlepage}

\tableofcontents

\section{Introduction}

The study of entanglement measures in many-body systems, such as the entanglement entropy \cite{bennet} and the logarithmic negativity \cite{Eisert, ZHSL, Eisert2, ple, erratumple, VW}, has led to many universal results, showing that entanglement encodes in a natural fashion fundamental aspects of quantum states at large scales \cite{CallanW94,HolzheyLW94,latorre1,Latorre2,Jin,Calabrese:2004eu,entropy,next,fractal,disco1,BCDLR,german1,german2,negativity1,negativity2,ourneg} (see also \cite{EERev1,specialissue,EERev2})).

Early results identified partition functions on certain Riemann surfaces in conformal field theory (CFT) as playing a fundamental role in the calculation of the von Neumann entanglement entropy of one-dimensional systems via the replica trick \cite{CallanW94,HolzheyLW94,Calabrese:2004eu}. In a modern language, such partition functions give the $n$th R\'enyi entropy. This concept was later generalised to massive quantum field theory (QFT) \cite{entropy}, where branch-point twist fields were identified, twist fields associated with cyclic permutation symmetries of models composed of $n$ independent copies. It is known that permutation twist fields of the $n$-copy replica model generate the Riemann surface connectivity \cite{kniz,orbifold}. The same idea also holds in spin chains and quantum lattices of any dimensionality \cite{permutation}, where, likewise, ``permutation twists" are involved, which are products of permutation operators on strings or higher-dimensional regions in the chain or lattice. Similar ideas can be used to evaluate the logarithmic negativity \cite{negativity1,negativity2}.

Partition functions on Riemann surfaces and branch-point twist fields have been used to obtain many results concerning the entanglement structure of vacuum states. Recently, attention has been given to the entanglement contribution of excitations above the vacuum, the increment of entanglement, as a probe for the nature of quantum excitations. This was first investigated in low-lying excitations of CFT \cite{german1,german2}. Excitations in massive QFT are generically believed to be of quite a different nature to low-lying states in CFT, having quasiparticle properties. In the first parts of this series of papers \cite{the4ofus1,the4ofus2,the4ofus3}, we argued that the entanglement offers a clear indication of the quasiparticle nature of massive excitations and of excitations with small de Broglie's wavelengths.

In excited states formed of a finite number of quasiparticle excitations, in the limit where the system's and regions' volumes are large, the increment of R\'enyi entanglement entropies due to the quasiparticles equates the entanglement entropy of certain ``quasiparticle qubit states", where qubits associated to the interior and exterior of the entanglement regions represent the presence or not of quasiparticles there, and amplitudes, their uniform distribution in space. This was proven  \cite{the4ofus2} for connected entanglement regions in the one-dimensional relativistic massive free boson and in the free Majorana fermion, using the form factor expansions of branch-point twist fields developed in \cite{entropy} and the finite-volume form factor theory developed in \cite{PT1,PT2}. It was also numerically verified in higher dimensions and shown in certain states of interacting models \cite{the4ofus1}. The result is thus expected to be quite general. The idea of the result -- identifying entanglement measure's increments with that of qubit states -- was also shown to hold for the ``replica logarithmic negativity" in the one-dimensional massive free boson and for the R\'enyi entropies of multiple disconnected regions \cite{the4ofus3}, again from a form factor analysis.

The goal of this paper is twofold. First, we show that the replica logarithmic negativity and R\'enyi entanglement entropy in quasiparticle qubit states are equal to {\em partition functions}, or generating functions, of certain families of graphs. These are weighted sums of graphs satisfying certain conditions, related to the {\em connectivity of the (abstract) manifold induced by the permutation-twist representation of the entanglement measures}. This works for arbitrary combinations of permutation twists, which might not have an immediate entanglement-measure interpretation.

Second, we show that the results of \cite{the4ofus1,the4ofus2,the4ofus3} for the replica logarithmic negativity and R\'enyi entanglement entropy are valid in free bosonic quantum field theory of {\em any dimension}, and with regions of {\em any shape and connectivity}. The proof is based on the result on graph partition functions, and uses a very different approach from that of form factors. Instead, it uses the expression of many-particle excited states in terms of local operators, and the basic exchange relations of permutation twists and clustering properties. Again, the proof makes a number of generalisations immediate, for instance to other combinations of permutation twists and perhaps to their descendants, to other quasiparticle excitations such as the particle and hole excitations above thermal or generalised Gibbs ensembles \cite{Essler2016} as considered in free models in \cite{DfiniteTff2005,DfiniteTff2007,ChenDoyon2014}, and, potentially, to interacting integrable models.

For completeness, we provide here the general statement concerning the graph partition functions. We consider the graph partition function associated to a certain combination of permutation twists in the system's manifold $\frak M$, and to a certain set of particles' momenta. For $m,m'\in\{1,\ldots,n\}$, we denote by $\frak R_{m,m'}\subset \frak M$ the total region where the permutations connect copy $m$ to $m'$. In the expressions of the qubit states, the volumes are only involved through the ratios of volumes $R_{m,m'}={\rm Vol}(\frak R_{m,m'})/{\rm Vol}(\frak M)$ via the uniform-distribution interpretation, see for instance \eqref{Psi} below. The graphs satisfy the following rules.
\bi
\item The graphs are composed of two disjoint finite sets of vertices of equal cardinality.
\item Each vertex is characterised by a copy label and a particle label, each copy being represented an equal number of times in both sets, and each particle being an equal number of times in each copy.
\item Each edge of the graph connects one vertex in a set to one in the other. Therefore there is no link between vertices in the same set.
\item All vertices are connected exactly once. Therefore there is no unpaired vertex.
\item Only vertices with labels of particles which have equal momenta can be connected.
\item Every edge connecting copy $m$ to $m'$ contributes to the evaluation of a graph $g$ a factor $R_{m,m'}$.
\ei
 \begin{figure}[h!]
 \begin{center} 
 \includegraphics[width=12cm]{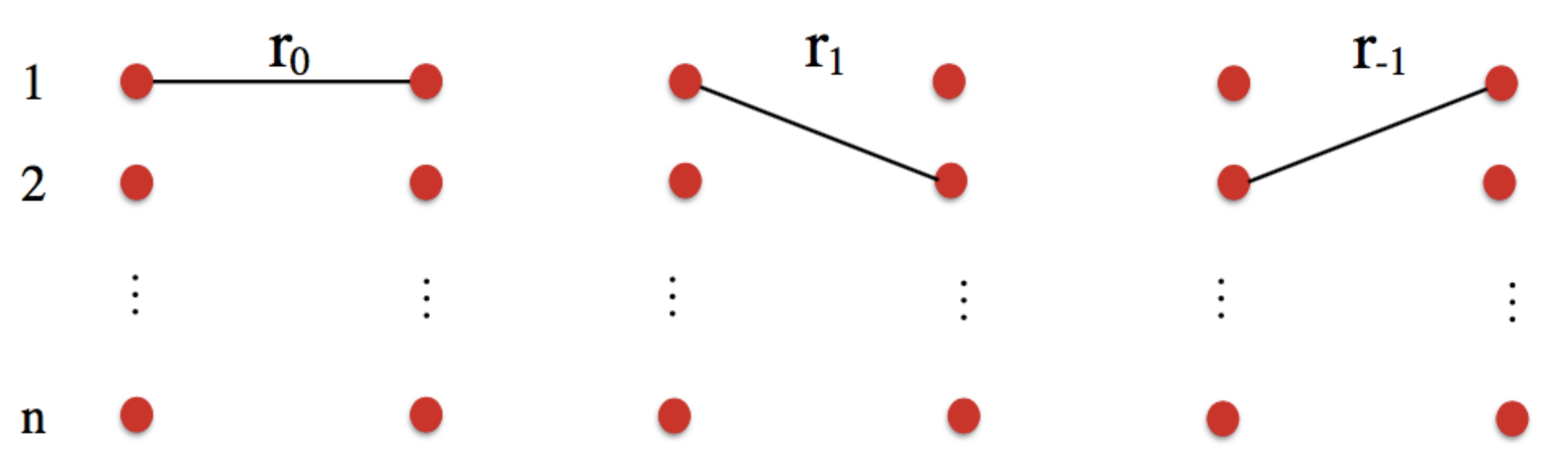} 
  \end{center} 
 \caption{The three building blocks for all connection rules in $\mathsf{G}_{1,n}$. The dots represent excitations. The connectivity of the particular entanglement measures considered dictates that these can only be connected horizontally or diagonally to another dot in the previous or the next row.  In a generic graph (for $k=1$), the number of dots in each column is $n$ and each dot on the left must be connected to a single dot on the right. Since only three types of link exist, this restricts the number of contributing graphs. The rules are analogous for $\mathsf{G}_{k,n}$, but in this case we have $k$ copies of this structure which we may represent by having dots of different colours. Then each dot on the left is connected to a dot on the right corresponding to the same copy number but possibly different $k$ (colour), see e.g. Fig.~\ref{graphsk2}.}
 \vspace{1cm}
 \label{negative1} 
 \end{figure}

That is, the graph partition function is
\beq\label{partitionfunctiongeneral}
	\sum_{g} \prod_{m,m'} (R_{m,m'})^{N_{m,m'}(g)}\,,
\eeq
where the power $N_{m,m'}(g)$ is the number of edges connecting copies $m$ to $m'$ in the graph $g$. In the present paper, for clarity we concentrate on entanglement measures, so that only cyclic permutations are involved, with $m=m'+\ell$ and $\ell\in\{1,0,-1\}$; however the above general statement will be clear from the proofs.

The paper is organised as follows. In section \ref{sectgraph} we state and prove the connection between replica negativity and R\'enyi entanglement entropy of quasiparticle qubit states and graph partition functions. In section \ref{sectqft} we show the general result in a free bosonic quantum field theory of arbitrary dimension. We conclude, and briefly discuss some generalisations, in section \ref{sectconclu}. In appendix \ref{appgraphs} we analyse some examples of the replica logarithmic negativity and the corresponding graph representation. In appendix \ref{sec:recur} we reproduce the analytic result of the single particle replica negativity presented in \cite{the4ofus3} through a recursion relation of the graph partition function.

\section{Graph partition functions for qubit entanglement}\label{sectgraph}

\subsection{Definitions and main statement}

\subsubsection{Graphs}
\label{sssectgraphs}

Consider the family of graphs described above \eqref{partitionfunctiongeneral} in the introduction. As mentioned, we specialise it to the cases which are of immediate use to the evaluation of the replica logarithmic negativity and the R\'enyi entanglement entropy, giving a precise definition for these cases. Throughout, for every $m\in\N$ we denote $I_m = \{1,\ldots,m\}$.

Let $k\in\N$ and $n\in\N$. With respect to the description above  \eqref{partitionfunctiongeneral}, $k$ is the number of particles, assuming they all have equal momenta, and $n$ is the number of copies. Consider the set ${\mathsf G}_{k,n}$ of all graphs as follows. They are formed by the $2kn$ vertices $\mathsf{V} = \{V_{j,m}^\ep:j\in I_k,\;m\in I_n,\;\ep\in\{\rl,\rr\}\}$ -- two sets, ``left" and ``right", of $kn$ vertices -- and $kn$  edges. The edges join left to right vertices, and are within the set $\mathsf E = {\mathsf E}_1\cup{\mathsf E}_0\cup{\mathsf E}_{-1}$ formed by the union of ${\mathsf E}_\ell = \{(V_{j,m}^\rl,V_{j',m+\ell}^\rr):j,j'\in I_k,\;m\in I_n\}$ (under the identification $V_{j,n+1}^{\rl,\rr} = V_{j,1}^{\rl,\rr}$, $V_{j,0}^{\rl,\rr} = V_{j,n}^{\rl,\rr}$), with the rule that each vertex must be attached to one and only one edge. Note that if $n=2$, then $\mathsf E_1 = \mathsf E_{-1}$, and thus in this case $\mathsf E = {\mathsf E}_1\cup{\mathsf E}_0$. For a graph $g\in{\mathsf G}_{k,n}$, denote by $N_\ell(g),\,\ell\in\{1,0,-1\}$ the number of edges of the graph that are in ${\mathsf E}_\ell$. We define the following polynomial in three variables $r_1,r_0,r_{-1}$, the {\em partition function of $\mathsf{G}_{k,n}$},
\beq\label{poly}
	p_{k,n}(r_1,r_0,r_{-1}) = \sum_{g\in{\mathsf G}_{k,n}} \cdot \lt\{
	\ba{ll}\displaystyle \prod_{\ell\in\{1,0,-1\}} r_\ell^{N_\ell(g)} & (n>2) \\
	\displaystyle r_0^{N_0(g)}(r_1+r_{-1})^{N_1(g)} & (n=2).
	\ea\rt.
\eeq
Note that for $n=2$, it is the sum $r_1+r_{-1}$ that is raised to the power of the number of edges in $\mathsf E_1=\mathsf E_{-1}$.

Below we make the precise connection with entanglement measures and permutation twists. Let us mention already that \eqref{poly} agrees with \eqref{partitionfunctiongeneral}, if the permutation-twist configuration is such that on a region of volume ratio $r_1$, every copy $m$ permutes to $m+1\;{\rm mod}\;n$, on a distinct region of volume ratio $r_0$, every copy $m$ permutes to $m$ (identity element), and on a still distinct region of volume ratio $r_{-1}$, every copy $m$ permutes to $m-1\;{\rm mod}\;n$. Making the connection with  \eqref{partitionfunctiongeneral}, with $n>2$ we have $R_{m,m+\ell} = r_\ell$ with $n=2$ we have $R_{1,1} = r_0$, $R_{1,2} = R_{2,1} = r_1+r_{-1}$, and in all cases $N_\ell(g) = \sum_m N_{m,m+\ell}(g)$.

We will also consider the subset $\mathsf{G}_{k,n}' \subset \mathsf{G}_{k,n}$, composed of all graphs where all edges lie within the set ${\mathsf E}_1\cup{\mathsf E}_0$, with labels $\ell=0,1$. This is a strict subset only if $n>2$. Its partition function is the following polynomial in two variables $r_1,r_0$:
\beq\label{poly2}
	p_{k,n}(r_1,r_0,0) = \sum_{g\in{\mathsf G}_{k,n}'} \prod_{\ell\in\{1,0\}} r_\ell^{N_\ell(g)}\,.
\eeq

\subsubsection{Replica negativity and entanglement entropy}

We now define more precisely the entanglement measures we consider.

Consider the Hilbert space ${\cal H}=L^2(\N_0^3)$ (where $\N_0 = \{0,1,2,\ldots\}$). Let us consider the orthonormal basis $\{|\underline{k}\ket:\underline{k} = (k_{1},k_0,k_{-1})\in \N_0^3\}$. This orthonormal basis naturally extracts the structure ${\cal H} = {\cal H}_1\otimes {\cal H}_0\otimes {\cal H}_{-1}$, with ${\cal H}_\ell\simeq L^2(\N_0)$, $\ell\in\{1,0,-1\}$ all isomorphic to each other (and to ${\cal H}$), whose orthonormal bases are naturally chosen as $\{|k_\ell\ket:k_\ell\in \N_0\}$. On the subspace ${\cal H}_{1,-1}= {\cal H}_1\otimes {\cal H}_{-1}$, we define as usual the partial transpose of an operator $A\in{\rm End}({\cal H}_1\otimes {\cal H}_{-1})$ by $\bra k_1,k_{-1}|A^{{\rm T}_{-1}}|k_1',k_{-1}'\ket = \bra k_1,k_{-1}'|A|k_1',k_{-1}\ket$. On ${\cal H}$, we also define, as usual, the partial trace $\Tr_{{\cal H}_0}:{\cal H} \to{\cal H}_{1,-1}$ of a trace-class operator $B\in{\rm End}({\cal H})$ by $\bra k_1,k_{-1}|\Tr_{{\cal H}_0}( B )|k_1',k_{-1}'\ket= \sum_{k_0=0}^\infty \bra k_1,k_0,k_{-1}| B |k_1',k_0,k_{-1}'\ket$.

Using these, the ``$n$-replica logarithmic negativity" $\mathcal{E}_n^{|\psi\ket}$ of tensor factors ${\cal H}_1$ and ${\cal H}_{-1}$  in the state $|\psi\ket\in{\cal H}$ (with $\bra\psi|\psi\ket=1$) is defined as follows:
\beq
	\exp\big[\mathcal{E}_n^{|\psi\ket}\big] = 
	\Tr_{{\cal H}_{1,-1}} \Big[\Big(\big(\Tr_{{\cal H}_0} (|\psi\ket\bra\psi|)\big)^{{\rm T}_{-1}}\Big)^n\Big]\,.
\eeq
The  logarithmic negativity provides a partial measure of the entanglement between ${\cal H}_1$ and ${\cal H}_{-1}$ \cite{Eisert, ZHSL, Eisert2, ple, erratumple, VW} and its replica version was first proposed in \cite{negativity1,negativity2}. The logarithmic negativity can be obtained as the unique analytic continuation of $\mathcal{E}_{2m}^{|\psi\ket}$ from integer values of $m$ to the value $1/2$, under appropriate specifications on its analytic structure as a function of $m$.

The R\'enyi entanglement entropy can be defined as a special case of the $n$-replica negativity. The R\'enyi entanglement entropy between tensor factors ${\cal H}_1$ and ${\cal H}_0$ in a vector $|\psi\ket' \in {\cal H}_1\otimes {\cal H}_0$ is simply obtained using the replica negativity of the vector $|\psi\ket'\otimes |0\ket$ (or any such factorised vector in ${\cal H}$) as follows:
\beq\label{SE}
	S_n^{|\psi\ket'} = \frc{\mathcal{E}^{
	|\psi\ket'\otimes |0\ket}_n}{1-n}\,.
\eeq
In this factorised case, the trace on tensor factor ${\cal H}_{-1}$ is trivial, and we obtain
\beq
	\exp\big[(1-n)S_n^{|\psi\ket'} \big] = 
	\Tr_{{\cal H}_{1}} \Big[\big(\Tr_{{\cal H}_0} (|\psi\ket'\ '\bra\psi|)\big)^n\Big]\,.
\eeq

\subsubsection{Qubit states}

We next specify the specific states in ${\cal H}$ whose entanglement measures give graph partition functions.

For $k\in\N$ and $r_1,r_0,r_{-1}\in[0,1]$ with $r_1+r_0+r_{-1}=1$, we define the vector
\beq\label{Psi}
	|\Psi_k(r_1,r_0,r_{-1})\ket = \sum_{\underline{k}=(k_1,k_0,k_{-1})\in \N_0^3\atop k_1+k_0+k_{-1}=k}
	\sqrt{\frc{ k! r_1^{k_1} r_0^{k_0} r_{-1}^{k_{-1}}}{k_1! k_0! k_{-1}!}}|\underline{k}\ket \in{\cal H}\,.
\eeq
Note that this vector is normalised:
\beq
	\bra\Psi_k(r_1,r_0,r_{-1})|\Psi_k(r_1,r_0,r_{-1})\ket = \sum_{(k_1,k_0,k_{-1})\in \N_0^3\atop k_1+k_0+k_{-1}=k}
	\frc{ k! r_1^{k_1} r_0^{k_0} r_{-1}^{k_{-1}}}{k_1! k_0! k_{-1}!}
	= (r_1+r_0+r_{-1})^k = 1\,.
\eeq

This vector represents the ``qubit state" for a flat distribution of $k$ indistinguishable particles amongst three complementary regions labelled by $\ell \in \{1,0,-1\}$, which have lengths $r_\ell$ adding to 1. The vector $|\underline{k}\ket$ is associated with the presence of $k_\ell$ particles in region $\ell$, and the square of the coefficient, the number $\frc{ k! r_1^{k_1} r_0^{k_0} r_{-1}^{k_{-1}}}{k_1! k_0! k_{-1}!}$, with $k_1+k_0+k_{-1}=k$, is the associated probability that this configuration occurs if we were to place randomly and independently, with uniform distribution, $k$ particles on the interval $[0,1]$ covered by three non-intersecting subintervals of lengths $r_1,r_0,r_{-1}$.

In the case where $r_{-1}=0$, then the distribution of the particles is over two non-intersecting regions labelled $\ell\in\{1,0\}$. The resulting vector is of the form
\beq\label{psiSE}
	|\Psi_k(r_1,r_0,0)\ket = 
	|\Psi_k(r_1,r_{0})\ket'\otimes |0\ket\,.
\eeq

\subsubsection{Theorems}

In \cite{the4ofus3} the following formulae for the $n$-replica logarithmic  negativity and the R\'enyi entropy of the state $|\Psi_k(r_1,r_0,r_{-1})\ket$ were proven, starting with the state (\ref{Psi}):
\begin{theorem} \label{thformula} Let $k\in\N$, $n\in\N$ and $r_1,r_0,r_{-1}\in[0,1]$ with $r_1+r_0+r_{-1}=1$. Then
\beq\label{formula}
	\exp\big[\mathcal{E}_n^{|\Psi_k(r_1,r_0,r_{-1})\ket}\big]
	 =
	\sum_{p=-k}^k \sum_{\sigma=\max(0,-np)}^{[\frac{n}{2}(k-p)]} {\mathcal{A}}_{p,\sigma} r_1^{np+\sigma} r_0^{n(k-p)-2\sigma} r_{-1}^{\sigma}\,.
\eeq
The coefficients ${\mathcal{A}}_{p,\sigma}$ are
\beq
{\mathcal{A}}_{p,\sigma}=\sum_{\{k_1,\ldots,k_n\}\in P_n(\sigma)} \prod_{j=1}^n \frac{k!}{(p+k_j)!(k-p-k_{j+1\;{\rm mod}\;n}-k_j)! k_{j+1\;{\rm mod}\;n}!}\,,
\label{pro}
\eeq
where $P_n(\sigma)$ represents the set of integer partitions of $\sigma$ into $n$ non-negative parts, and by convention the product is set to zero whenever any argument of the factorials is negative. Further,
\beq\label{formula2}
	\exp\big[(1-n)S_n^{|\Psi_k(r_1,r_0)\ket'}\big]
	= \sum_{j=0}^k\lt(\frc{k!}{j!(k-j)!}r_1^j r_0^{k-j}\rt)^n\,.
\eeq
\end{theorem}
Note that if $n$ is even, $n=2m$ with $m\in\N$, then (\ref{formula}) can be written in a more symmetric fashion,
\beq
	\exp\big[\mathcal{E}_{2m}^{|\Psi_k(r_1,r_0,r_{-1})\ket}\big] =
	\sum_{p=-k}^k \sum_{\sigma=|mp|}^{mk}
	A_{p,\sigma-mp} r_1^{\sigma+mp} r_0^{2(mk-\sigma)} r_{-1}^{\sigma-mp}\,.
\eeq
Note also that the case $r_{-1}=0$ of \eqref{formula} immediately gives \eqref{formula2}.

The main goal of this section is to prove the following theorem:
 The $n$-replica negativity of the state $|\Psi_k(r_1,r_0,r_{-1})\ket$ is proportional to the polynomial \eqref{poly} up to a simple numerical factor; entanglement is thus related to the combinatoric problem of counting the graphs ${\mathsf G}_{k,n}$, and this counting problem leads to the explicit formula \eqref{formula}. 
The precise mathematical statement is:
\begin{theorem}\label{thmain} Let $k\in\N$, $n\in\N$ and $r_1,r_0,r_{-1}\in[0,1]$ with $r_1+r_0+r_{-1}=1$. Then
\beq\label{formg}
	\exp\big[\mathcal{E}_n^{|\Psi_k(r_1,r_0,r_{-1})\ket}\big] = \frc{p_{k,n}(r_1,r_0,r_{-1})}{(k!)^n}\,,
\eeq
and
\beq\label{formg2}
	\exp\big[(1-n)S_n^{|\Psi_k(r_1,r_0)\ket'}\big]
	= \frc{p_{k,n}(r_1,r_0,0)}{(k!)^n}\,.
\eeq
\end{theorem}
We note that, again, \eqref{formg2} immediately follows from \eqref{formg} using \eqref{SE} and \eqref{psiSE}, hence below we only prove \eqref{formg}.

\subsection{Proof}

In this section we will use the notation $k_\ell^{(j)}$ where $j$ is the replica number, and for notational convenience, we make the identifications
\beq\label{id}
	k_\ell^{(0)} \equiv k_\ell^{(n)},\quad k_\ell^{(n+1)} \equiv k_\ell^{(1)}\,.
\eeq
In order to show Theorem \ref{thmain}, we show two lemmas.

\subsubsection{Permutation twists}\label{permutation}

First, we re-write the $n$-replica logarithmic negativity in terms of a quantum average, in the state represented by $|\psi\ket^{\otimes n}$ on the $n$-replica Hilbert space ${\cal H}^{\otimes n}$, of a particular product of permutation operators, permuting the copies and acting on the individual tensor factors ${\cal H}_\ell$ -- these are the permutation twists. For the entanglement measures, it is sufficient to consider cyclic permutations. We define the operators $\mathbf{P}_\ell^\pm$ for $\ell\in\{1,0,-1\}$, acting on ${\cal H}^{\otimes n}$ as
\beq
	 \mathbf{P}_\ell^\ep\, \bigotimes_{j=1}^n |k_1^{(j)},k_0^{(j)},k_{-1}^{(j)}\ket
	 = 
	\bigotimes_{j=1}^n |k_1^{(j-\ep\delta_{1,\ell})},k_0^{(j-\ep\delta_{0,\ell})},k_{-1}^{(j-\ep\delta_{-1,\ell})}\ket
	,\quad \ep\in\{+,-\}\,,
\eeq
where $\bigotimes_{j=1}^n|\underline{k}^{(j)}\ket = |\underline{k}^{(1)};\ldots;\underline{k}^{(n)}\ket\in{\cal H}^{\otimes n}$, and with the identifications \eqref{id}. These operators perform cyclic permutations of the $n$ copies of the individual tensor factors ${\cal H}_\ell$; with $\ep=+$ ($-$), they shift them rightwards (leftwards). With the notation that the operator $\Or^{(j)}_\ell$  acts nontrivially only on the $j^{\rm th}$ tensor factor of ${\cal H}^{\otimes n}$, and on this factor, nontrivially only on the tensor factor ${\cal H}_\ell$ of ${\cal H}$, as $\Or\in{\rm End}({\cal H}_\ell)$, the permutation operators satisfy the {\em exchange relations}
\beq\label{exchange}
	\mathbf{P}^\ep_\ell \Or^{(j)}_{\ell'} = \lt\{\ba{ll}
	\Or^{(j+\ep)}_{\ell'}
	\mathbf{P}^\ep_\ell & (\ell=\ell') \\
	\Or^{(j)}_{\ell'} \mathbf{P}^\ep_\ell & (\ell\neq\ell')
	\ea\rt.\,.
\eeq
The following holds:
\begin{lemma} \label{lemtwist} Let $|\psi\ket\in{\cal H}$ with $\bra\psi|\psi\ket=1$. Then
\beq\label{negtwist}
	\exp\big[\mathcal{E}_n^{|\psi\ket}\big] =
	{}^{\otimes n}\bra\psi| \mathbf{P}_1^+ \mathbf{P}_{-1}^-|\psi\ket^{\otimes n}\,.
\eeq
Let $|\psi\ket'\in{\cal H}_1\otimes{\cal H}_0$. Then
\beq\label{formg22}
	\exp\big[(1-n)S_n^{|\psi\ket'}\big] =
	{}^{\otimes n}\,{}'\bra\psi| \mathbf{P}_1^+ |\psi\ket'{}^{\otimes n}\,,
\eeq
where by a slight abuse of notation, $\mathbf{P}_1^+$ is the natural restriction to ${\rm End}({\cal H}_1\otimes{\cal H}_0)$.

\end{lemma}

\proof The proof is obtained by a direct evaluation of both sides. From \cite[Eq 3.20]{the4ofus3}, writing $|\psi\ket = \sum_{\underline{k}\in\N_0^3} c_{\underline{k}}|\underline{k}\ket$, we have
\beq\label{neggen}
	\exp\big[\mathcal{E}_n^{|\psi\ket}\big] =
	\sum_{\{k_\ell^{(j)}\in\N_0:\atop \ell\in\{1,0,-1\},j\in I_n\}}
	\prod_{j=1}^{n} c_{k_1^{(j+1)},k_0^{(j)},k_{-1}^{(j)}}c_{k_1^{(j)},k_0^{(j)},k_{-1}^{(j+1)}}^*\,,
\eeq
with the convention \eqref{id}. This gives the left-hand side of \eqref{negtwist}. On the other hand, we have
\beq
	\mathbf{P}_{1}^+\mathbf{P}_{-1}^-|\psi\ket^{\otimes n}
	= \sum_{\{k_\ell^{(j)}\in\N_0:\atop \ell\in\{1,0,-1\},j\in I_n\}}
\prod_{j=1}^n
	c_{k_1^{(j)},k_0^{(j)},k_{-1}^{(j)}}\,
	\bigotimes_{j=1}^n |k_1^{(j-1)},k_0^{(j)},k_{-1}^{(j+1)}\ket\,,
\eeq
which gives rise to
\beq
	{}^{\otimes n}\bra \psi|\mathbf{P}_{1}^+\mathbf{P}_{-1}^-|\psi\ket^{\otimes n}
	= \sum_{\{k_\ell^{(j)}\in\N_0:\atop \ell\in\{1,0,-1\},j\in I_n\}}
	\prod_{j=1}^{n} c_{k_1^{(j+1)},k_0^{(j)},k_{-1}^{(j)}}c_{k_1^{(j)},k_0^{(j)},k_{-1}^{(j+1)}}^*\,,
\eeq
showing the first part of the lemma. The second part is obtained immediately by using \eqref{SE}.
\eproof

Lemma \ref{lemtwist} makes it clear that, despite the partial transpose used in the definition of the $n$-replica negativity, the result is invariant under unitary transformations of the individual tensor factors of ${\cal H}$ (this is a well-known fact):
\begin{corol} \label{corinv}The $n$-replica negativity is invariant under unitary transformations of its factors,  $\mathcal{E}_n^{|\psi\ket} =  \mathcal{E}_n^{U_{1}U_0 U_{-1}|\psi\ket}$ for any unitary $U_\ell$ acting nontrivially on ${\cal H}_\ell$.
\end{corol}

\subsubsection{Fock space representation}

Second, we establish using  \eqref{negtwist}, for rational values of $r_\ell$, a representation of the replica logarithmic negativity of the state $|\psi\ket=|\Psi_k(r_1,r_0,r_{-1})\ket$, as that of a new state in a Fock space representing particles on a chain. The Fock-space state is the $k$th power of a uniform sum, over all positions of the system, of position-labelled particle creation operators, representing the idea that particles are uniformly distributed in space. This makes the interpretation of the quasiparticle qubit state \eqref{Psi} clearer, and will directly lead, by the exchange relation \eqref{exchange} and Wick's theorem, to the graph partition functions \eqref{poly}.

Let $L\in\N$, set $X_{-2}=0$ and $X_1=L$, and let $X_{-1},X_0\in \{1,\ldots,L\}$ such that $X_{\ell-1}< X_\ell$ for $\ell=-1,0,1$. Consider the non-intersecting subsets $\frak R_\ell = \{X_{\ell-1}+1,\ldots,X_\ell\}$ for $\ell=-1,0,1$, which have cardinalities $L_\ell = |\frak R_\ell| = X_\ell-X_{\ell-1}$ summing to $L_1 + L_0 + L_{-1} = L$. Set
\beq
	r_\ell = \frc{L_\ell}{L},\quad \ell=-1,0,1.
\eeq
Construct the Fock space ${\cal F}$ with the canonical commutation relations for the operators $a_x,\,a_y\in{\rm End}({\cal F}),\;x,y\in\{1,\ldots,L\}$,
\beq
	[a_x,a_y^\dag] = \delta_{x,y},\quad
	[a_x,a_y] = [a_x^\dag,a_y^\dag]=0,\quad x,y\in\{1,\ldots,L\}\,,
\eeq
and with the vacuum $|0\dket$ satisfying $a_x|0\dket = 0\;\forall x$. This factorises as ${\cal F} = {\cal F}_1 \otimes {\cal F}_0\otimes {\cal F}_{-1}$ into Fock spaces ${\cal F}_\ell$ for the generators $\{a_x,a_x^\dag:x\in \frak{R}_{\ell}\}$. As any two countable-dimensional Hilbert spaces are isomorphic, we have isomorphisms ${\cal F}_\ell \simeq{\cal H}_\ell$ for $\ell=-1,0,1$, and therefore  ${\cal F}\simeq{\cal H}$. Let us denote one such isomorphism by $\phi_{\cal F}:{\cal F}\to {\cal H}$, with $\phi_{\cal F}({\cal F}_\ell) = {\cal H}_\ell$. We define the permutation twists
\beq\label{PF}
	\mathbf{P}_\ell^{\ep,{\cal F}} := \phi_{\cal F}^{-1}\circ\mathbf{P}_\ell^\ep\circ \phi_{\cal F}
\eeq
on ${\cal F}^{\otimes n}$. This acts, in the natural way, by permutation of the copies on the individual tensor factors ${\cal F}_\ell$, and is independent of the choice of $\phi_{\cal F}$. Then, following Lemma \ref{lemtwist}, for any $|\psi\dket\in{\cal F}$, we define the $n$-replica logarithmic negativity on ${\cal F}$ by\footnote{By a slight abuse of notation, we use the same symbol for the replica negativity, the difference being in the symbol used for the vector, which specifies the space in which it lies.} 
\beq
	\exp\big[\mathcal{E}_n^{|\psi\dket}\big]
	= {}^{\otimes n}\dbra\psi| \mathbf{P}_1^{+,{\cal F}} \mathbf{P}_{-1}^{-,{\cal F}}|\psi\dket^{\otimes n}\,.
\eeq

Let
\beq\label{nfp}
	|\Psi_k(r_1,r_0,r_{-1})\dket=
	\frc1{
	\sqrt{k!L^k}
	}
	\Bigg(\sum_{x\in \{1,\ldots,L\}} 
	a^\dag_{x}\Bigg)^k |0\dket \in {\cal F}\,.
\eeq
Writing $\sum_{x\in \{1,\ldots,L\}} = \sum_{x\in \mathfrak R_1} + \sum_{x\in \mathfrak R_2} + \sum_{x\in \mathfrak R_3}$, one can regroup the terms in the following way:
\beqa
	|\Psi_k(r_1,r_0,r_{-1})\dket&=&
	\frc1{
	\sqrt{k!L^k}
	}
	\sum_{k_1,k_0,k_{-1}\in \{0,\ldots,k\}\atop k_{1}+k_0+k_{-1}=k}
	\frc{k!}{k_1!k_0!k_{-1}!}
	\prod_{\ell\in\{1,0,-1\}}
	\Bigg(\sum_{x\in \frak{R}_\ell}
	a_x^\dag\Bigg)^{k_\ell}
	 |0\dket\n
	 &=&
	\sum_{k_1,k_0,k_{-1}\in \{0,\ldots,k\}\atop k_{1}+k_0+k_{-1}=k}
	\sqrt{\frc{k!r_1!r_2!r_3!}{k_1!k_0!k_{-1}!}}
	\prod_{\ell\in\{1,0,-1\}}
	\frc1{\sqrt{k_\ell!L_\ell^{k_\ell}}}
	\Bigg(\sum_{x\in \frak{R}_\ell}
	a_x^\dag\Bigg)^{k_\ell}
	 |0\dket \,. \label{rewriting}
\eeqa
We see that this has the structure of the vector $|\Psi_k(r_1,r_0,r_{-1})\ket$ defined in \eqref{Psi}. This allows us to show the following lemma, which makes the correspondence explicit. Via this correspondence, the nontrivial combinatoric factors in \eqref{rewriting} are seen explicitly to occur, using the expression \eqref{nfp}, from a uniform distribution of particles in space. This is what will lead, in paragraph \ref{paraproof}, to the proof of the relation with graph partition functions.
\begin{lemma}\label{lemnf}
\beq
	\mathcal{E}_n^{|\Psi_k(r_1,r_0,r_{-1})\ket} = \mathcal{E}_n^{|\Psi_k(r_1,r_0,r_{-1})\dket}.
\eeq
\end{lemma}
\proof In ${\cal F}_\ell$, construct the vectors
\beq
	|k_\ell\dket_\ell =
		\frc1{\sqrt{k_\ell!L_\ell^{k_\ell}}}
	\Bigg(\sum_{x\in \frak{R}_\ell}
	a_x^\dag\Bigg)^{k_\ell}
	 |0\dket\n	 
	=
	\sum_{x_1,\ldots,x_{k_\ell}\in \frak{R}_\ell}
	\frc{\Big(
	\prod_{j\in I_{k_\ell}}
	a^\dag_{x_j}\Big) |0\dket}{
	\sqrt{k_\ell!L_\ell^{k_\ell}
	}}\in {\cal F}_\ell\,.
\eeq
These are orthonormal, ${}_{\ell}\dbra k_\ell|k_\ell'\dket_{\ell}= \delta_{k_\ell,k_\ell'}$. Construct the following vectors in ${\cal F}$:
\beq
	|k_1,k_0,k_{-1}\dket = 
	|k_1\dket_{1}\otimes |k_0\dket_{0}
	\otimes |k_{-1}\dket_{{-1}} \in {\cal F}\,,
\eeq
which are also orthonormal
$\dbra k_1,k_0,k_{-1}|k_1',k_0',k_{-1}'\dket= \delta_{k_1,k_1'}\delta_{k_0,k_0'}\delta_{k_{-1},k_{-1}'}$.
Finally, consider the subspace
\beq
	{\cal V} = {\rm span}\big(|k_1,k_0,k_{-1}\dket:k_1,k_0,k_{-1}\in \N_0\big)\subset {\cal F}\,.
\eeq
Clearly, there is an isomorphism from ${\cal V}$ onto ${\cal H}$, which can be explicitly written as
\beq
	\ba{rcl}\phi_{\cal V}:\quad\qquad\qquad{\cal V}&\to&{\cal H}\\
	|k_1,k_0,k_{-1}\dket &\mapsto& |k_1,k_0,k_{-1}\ket\,.
	\ea
\eeq
We define $\mathbf{P}_\ell^{\ep,{\cal V}}$, the permutation twists on ${\cal V}$, in a similar way to \eqref{PF}, using, say, the isomorphism $\phi_{\cal V}$. Clearly, from this definition, $\mathbf{P}_{\ell}^{\ep,{\cal V}}|\psi\dket^{\otimes n} = \mathbf{P}_{\ell}^{\ep,{\cal F}}|\psi\dket^{\otimes n}$ for all $|\psi\dket \in{\cal V}$. Therefore,
\beqa
	\exp\big[{\cal E}_n^{|\psi\ket}\big]
	&=&{}^{\otimes n}\bra\psi| \mathbf{P}_1^{+} \mathbf{P}_{-1}^{-}|\psi\ket^{\otimes n}
	={}^{\otimes n}\bra\psi|\phi_{\cal V}\mathbf{P}_1^{+,{\cal V}} \mathbf{P}_{-1}^{-,{\cal V}}\phi_{\cal V}^{-1}|\psi\ket^{\otimes n}
	={}^{\otimes n}\bra\psi|\phi_{\cal V}\mathbf{P}_1^{+,{\cal F}} \mathbf{P}_{-1}^{-,{\cal F}}\phi_{\cal V}^{-1}|\psi\ket^{\otimes n}\n
	&=& \exp\big[{\cal E}_n^{|\psi\dket}\big],\qquad
	|\psi\dket = \phi_{\cal V}^{-1}|\psi\ket 
	\in {\cal V}\subset {\cal F}.
\eeqa
Finally, as is clear from \eqref{rewriting},
\beq
	\phi_{\cal V}^{-1}|\Psi_k(r_1,r_0,r_{-1})\ket
	= |\Psi_k(r_1,r_0,r_{-1})\dket.
\eeq
This shows the lemma. \eproof

\subsubsection{Proof of theorem 2.2}\label{paraproof}

We use Lemma \ref{lemnf} along with \eqref{nfp} and write
\beq
	\exp\big[\mathcal{E}_n^{|\Psi_k(r_1,r_0,r_{-1})\ket}\big]
	=
	{}^{\otimes n}\dbra\Psi_k(r_1,r_0,r_{-1}| \mathbf{P}_1^{+,{\cal F}} \mathbf{P}_{-1}^{-,{\cal F}}|\Psi_k(r_1,r_0,r_{-1}\dket^{\otimes n}\,.
\eeq
The right-hand side is, explicitly,
\beq
	\frc1{(k!)^nL^{kn}}
\Big(\sum_{y_1,\ldots,y_k\in I_L}\dbra0|\prod_{j\in I_k} a_{y_j}\Big)^{\otimes n}\;
	\mathbf{P}_1^{+,{\cal F}} \mathbf{P}_{-1}^{-,{\cal F}}\;
	\Big(\sum_{x_1,\ldots,x_k\in I_L}\prod_{i\in I_k} a_{x_i}^\dag |0\dket\Big)^{\otimes n}\,.
\eeq
Let us denote by $a_{y}^{p}$ and $\big[a_{x}^{m}\big]^\dag$ the annihilation and creation operators on copy $p,m\in\{1,\ldots,n\}$. Using the exchange relation \eqref{exchange} and the invariance of the state $|0\dket^{\otimes n}$ under permutations, we pass the creation operators to the left of the permutation operators and obtain
\beq
	\frc1{(k!)^nL^{kn}}
	\sum_{\{y_{j,p}\in I_L\}}
	\sum_{\{x_{i,m}\in I_L\}}
	\dbra 0|
	\prod_{j\in I_k,\,p\in I_n} 
	a_{y_{j,p}}^p
	\prod_{i\in I_k,\,m\in I_n} \Big[a_{x_{i,m}}^{m+\chi(x_{i,m}\in \frak R_{1})-\chi(x_{i,m}\in \frak R_{-1})}\Big]^\dag
	|0\dket\,,
\eeq
where $\chi({\rm c})$ is the indicator function for condition ${\rm c}$, that is 1 if $\rm c$ is true and 0 otherwise. 
We evaluate this expression by Wick's theorem. Accordingly, the quantity 
\beq\label{qtytoe}
	\sum_{\{y_{j,p}\in I_L\}}
	\sum_{\{x_{i,m}\in I_L\}}\dbra 0|
	\prod_{j\in I_k,\,p\in I_n} 
	a_{y_{j,p}}^p
	\prod_{i\in I_k,\,m\in I_n} \Big[a_{x_{i,m}}^{m+\chi(x_{i,m}\in \frak R_{1})-\chi(x_{i,m}\in \frak R_{-1})}\Big]^\dag
	|0\dket\,,
\eeq
is a sum of Wick terms, each term being a product of Wick contractions between creation and annihilation operators, which simply evaluates to 1. We organise this sum as follows. Recall subsection \ref{sssectgraphs} where we introduced the graphs in $\mathsf{G}_{k,n}$. We identify each pair $j,p$ of labels in the product $\prod_{j\in I_k,\,p\in I_n} 
	a_{y_{j,p}}^p$ with the vertex $V_{j,p}^{\rr}$ in $\mathsf{V}$; and likewise we identify each pair $i,m$ of labels in the product $\prod_{i\in I_k,\,m\in I_n} \Big[a_{x_{i,m}}^{m+\chi(x_{i,m}\in \frak R_{1})-\chi(x_{i,m}\in \frak R_{-1})}\Big]^\dag$ with the vertex $V_{j,p}^{\rl}$ in $\mathsf{V}$. We also identify each Wick contraction in a Wick term with an edge between these vertices. Thus each Wick term is unambiguously a graph with edges connecting vertices in $\mathsf{V}$. There is a contraction between $a_{y_{j,p}}^p$ and $\Big[a_{x_{i,m}}^{m+\chi(x_{i,m}\in \frak R_{-1})-\chi(x_{i,m}\in \frak R_{1})}\Big]^\dag$ if and only if $y_{j,p}=x_{i,m}$ and $p = m+\chi(x_{i,m}\in \frak R_{1})-\chi(x_{i,m}\in \frak R_{-1})$. Therefore, either $p=m$, or $p=m+1$, or $p=m-1$. There are no other contractions. Hence, each edge is in $\mathsf{E}_1\cup\mathsf{E}_0\cup\mathsf{E}_{-1}$ (and $p=m+\ell$ corresponds to an edge in $\mathsf{E}_\ell$), as illustrated by Fig.~1. Further, by Wick's theorem, every vertex is the end-point of one and only one edge. Therefore, each Wick term is identified with a graph in $\mathsf{G}_{n,k}$, and each such term evaluates to 1.

We now need to count how many times  $N(g)$ a given graph $g\in\mathsf{G}_{n,k}$ occurs in the sum of Wick terms; then the result is written as
\[
	\sum_{\{y_{j,p}\in I_L\}}
	\sum_{\{x_{i,m}\in I_L\}}\dbra 0|
	\prod_{j\in I_k,\,p\in I_n} 
	a_{y_{j,p}}^p
	\prod_{i\in I_k,\,m\in I_n} \Big[a_{x_{i,m}}^{m+\chi(x_{i,m}\in \frak R_{-1})-\chi(x_{i,m}\in \frak R_{1})}\Big]^\dag
	|0\dket
	= \sum_{g\in\mathsf{G}_{n,k}} N(g)\,.
\]
For every edge $(V_{i,m}^{\rl},V_{j,p}^{\rr})$, there is a factor coming from the sum over the possible values of $y_{j,p}$ and $x_{i,m}$ leading to this edge. Because of the condition $y_{j,p}=x_{i,m}$, we only need to consider one sum. Because of the condition $p = m+\chi(x_{i,m}\in \frak R_{1})-\chi(x_{i,m}\in \frak R_{-1})$, the values of $y_{j,p}$ leading to this edge are all values $y_{j,p}\in \frak R_0$ if $p=m$, all values $y_{j,p}\in \frak R_1$ if $p=m+1$, and all values $y_{j,p}\in \frak R_{-1}$ if $p=m-1$. Thus, if $n>2$, for each edge in $\mathsf{E}_\ell$, there is a factor $|\frak R_\ell|= L_\ell$; and if $n=2$, then for each edge in $\mathsf{E}_1 = \mathsf E_{-1}$, there is a factor $L_1+L_{-1}$. Since there are exactly $kn$ edges, we obtain, for $n>2$,
\[
	\sum_{\{y_{j,p}\in I_L\}}
	\sum_{\{x_{i,m}\in I_L\}}\dbra 0|
	\prod_{j\in I_k,\,p\in I_n} 
	a_{y_{j,p}}^p
	\prod_{i\in I_k,\,m\in I_n} \Big[a_{x_{i,m}}^{m+\chi(x_{i,m}\in \frak R_{-1})-\chi(x_{i,m}\in \frak R_{1})}\Big]^\dag
	|0\dket
	= L^{kn}\sum_{g\in\mathsf{G}_{n,k}} r_\ell^{N_\ell(g)}\,,
\]
where we recall $r_\ell = L_\ell/L$ and $N_\ell(g)$ is the number of edges in $g$ that lie in $\mathsf{E}_\ell$; and
thus
\beq
	\exp\big[\mathcal{E}_n^{|\Psi_k(r_1,r_0,r_{-1})\ket}\big] =
	\frc1{(k!)^n}\sum_{g\in\mathsf{G}_{k,n}} r_\ell^{N_\ell(g)} = \frc{p_{k,n}(r_1,r_0,r_{-1})}{(k!)^n}\,.
\eeq
For $n=2$, a similar argument leads again to $\frc{p_{k,2}(r_1,r_0,r_{-1})}{(k!)^2}$.
\eproof

\section{Entanglement of particle excitations in free bosonic quantum field theory} \label{sectqft}

\subsection{Main statement}

Consider the massive free boson of mass $m$ on the hypertorus ${\frak M} = \times_{j=1}^d [0,L_j]\subset \R^d$ of dimension $d\geq 1$ (space-time having dimension $d+1$). For simplicity we assume that there is some UV regularisation, for instance a harmonic lattice. We will not need to specify any particular regularisation, since the derivation holds true as long as the general properties stated below remain valid. This in fact serves to illustrate the generality of the method, beyond the realm of UV-completed field theory.

We denote by $\lambda{\frak M}$ the hypertorus scaled by the factor $\lambda>1$. Consider two non-intersecting open subsets ${\frak R}_1,\;{\frak R}_{-1}\subset {\frak M}$, of any connectivity, and similarly denote by $\lambda{\frak R}_{1}$ and $\lambda{\frak R}_{1}$ the scaled subsets of $\lambda{\frak M}$. For simplicity of the argument, we assume that the regions ${\frak R}_1,\;{\frak R}_{-1}$ have piecewise smooth boundaries.

It is a simple matter to construct the vacuum state $|\vac\ket$ and multi-particle excited states $|\bm p_1 ,\ldots \bm p_k\ket$ in this theory via the Fock space ${\cal H}$ over the canonical algebra of annihilation and creation operators $A_{\bm{p}}$ and $A^\dag_{\bm{p}}$ at momenta $\bm p \in \Lambda_d\subset \R^d$. The momenta are quantised to the square lattice $\Lambda_d = \times_{j=1}^d (2\pi L_j^{-1} \Z)$, and on $\lambda {\cal M}$ they are quantised to $\lambda^{-1}\Lambda_d$. We have
\beq
	[A_{\bm p},A^\dag_{\bm p'}] = \delta_{\bm p,\bm p'},\quad A_{\bm p}|\vac \ket = 0,\quad
	|\bm p_1 ,\ldots \bm p_k\ket
	= A^\dag_{\bm p_1}\cdots A^\dag_{\bm p_k}|\vac\ket\,.
\eeq
These operators can be written in terms of the Klein-Gordon field $\Phi(\bm x)$ and its canonical conjugate $\Pi(\bm x)$ as
\beq\label{Ap}
	A_{\bm p} = \frc1{\sqrt{{\rm Vol}(\frak M)}}\int_{{\frak M}} \dd^d \bm x\,
	\re^{-\ri \bm p\cdot\bm x}\Or_{\bm p}(\bm x)\,,
\eeq
where
\beq
	\Or_{\bm p}(\bm x) = \frc{E_{\bm p}\Phi(\bm x) + \ri\Pi(\bm x)}{\sqrt{2E_{\bm p}}}\,,
\eeq
with $E_{\bm p}$ the energy. In the relativistic boson, it obeys the relativistic dispersion relation $E_{\bm p} = \sqrt{m^2+\bm p^2}$, but this is not necessary for the proof; other dispersion relations, such as that from the harmonic lattice, can be used. All states are normalised to 1.

We are interested in the increment of entanglement between two regions due to the presence of a finite number of particles. Thus we would like to evaluate the difference of replica logarithmic negativities, and of R\'enyi entropies, between a $k$-particle state and the vacuum $|\vac\ket$. The clearest way to define the replica entanglement negativity and the R\'enyi entanglement entropy in QFT is to use their general expressions in terms of permutation twists, shown in Lemma \ref{lemtwist}. We simply identify the tensor factors ${\cal H}_\ell$ with the spaces of field configurations on the regions $ {\mathfrak R}_\ell$, $\ell\in\{1,-1\}$; in general we will denote by $\mathbf{P}^\ep(\frak R)$ the permutation twists associated to the tensor factors of field configurations supported on $\frak R$. As we have assumed that there is some UV regularisation, no divergence occurs in averages of such permutation twists. In 1+1-dimensional quantum field theory, $\mathbf{P}^\ep(\frak R)$ is the product of appropriate {\em branch-point twist fields} positioned at the boundary points of $\frak R$ \cite{Calabrese:2004eu,entropy}.

Consider a set of momenta $\mathsf{p} = \{\bm p_1,\ldots,\bm p_k\}$ with $\bm p_j\in\Lambda_d$. We wish to evaluate, in an appropriate limit, the following replica logarithmic negativity and R\'enyi entropy increments,
\beq\label{QFTE}
	\exp\big[\Delta\mathcal{E}_n^{|\mathsf p\ket}({\frak R}_1,{\frak R}_{-1};{\frak M})\big]
	= \frc{
	{}^{\otimes n}\bra\mathsf{p}| \mathbf{P}^+(\frak R_1) \mathbf{P}^-(\frak R_{-1})|\mathsf{p}\ket^{\otimes n}}{
	{}^{\otimes n}\bra\mathsf{p}|\mathsf{p}\ket^{\otimes n}\ 
	{}^{\otimes n}\bra\vac| \mathbf{P}^+(\frak R_1) \mathbf{P}^-(\frak R_{-1})|\vac\ket^{\otimes n}}\,,
\eeq
and
\beq
	\exp\big[(1-n)\Delta S_n^{|\mathsf p\ket}({\frak R}_1;{\frak M})\big]
	= \frc{
	{}^{\otimes n}\bra\mathsf{p}| \mathbf{P}^+(\frak R_1)|\mathsf{p}\ket^{\otimes n}}{
	{}^{\otimes n}\bra\mathsf{p}|\mathsf{p}\ket^{\otimes n}\ 
	{}^{\otimes n}\bra\vac| \mathbf{P}^+(\frak R_1)|\vac\ket^{\otimes n}}\,,
\eeq
respectively.

In order to study these objects, we need some facts about the permutation twists in the $n$-copy massive free boson.

The main properties of the operators $\mathbf{P}^\ep(\frak R_\ell)$ are the exchange relations \eqref{exchange}, which here can be written
\beq\label{QFTexc}
	\mathbf{P}^\ep(\frak R_\ell) \Or^{(j)}(\bm x) = \lt\{\ba{ll}
	\Or^{(j+\ep)}(\bm x)
	\mathbf{P}^\ep(\frak R_\ell) & (\bm x\in\frak R_\ell) \\
	\Or^{(j)}(\bm x) \mathbf{P}^\ep(\frak R_\ell) & (\bm x\in\frak M\setminus \frak R_\ell)\,.
	\ea\rt.
\eeq
In this notation, $\Or^{(j)}(\bm x)$ is defined by acting nontrivially only on the $j^{\rm th}$ tensor factor of ${\cal H}^{\otimes n}$, and on this factor, it acts as the local operator $\Or(\bm x)\in{\rm End}(\cal H)$ positioned at $\bm x\in\frak M$. The set of fields formed by $\Or^{(j)}(\bm x)$ with $\Or(\bm x)\in\{{\bf 1},\,\Or_{\bm p}(\bm x),\Or_{\bm p}(\bm x)^\dag:{\bm p}\in\Lambda_d,\,\bm x\in\frak M\}$ and their products spans a dense subset of ${\rm End}({\cal H})^{\otimes n}$.

Because the theory has nonzero mass, all correlation functions of local operators factorise into products of correlation functions exponentially fast with the distance between operators. Something similar is expected (and verified in one dimension) to hold for the permutation operator. Let us express some of these {\em clustering properties} more precisely, in a way that is convenient for the proof below.

Consider the normalised correlation function
\[
	{}_{\mathbf P}\bra
	\Or_{\bm p_1}^{(j_1)}(\bm x_1)
	\cdots
	\Or_{\bm p_k}^{(j_k)}(\bm x_k)
	\ket_{\mathbf P}
	= \frc{{}^{\otimes n}\bra\vac| 
	\Or_{\bm p_1}^{(j_1)}(\bm x_1)
	\cdots
	\Or_{\bm p_k}^{(j_k)}(\bm x_k)
	\mathbf{P}^+(\frak R_1) \mathbf{P}^-(\frak R_{-1})|\vac\ket^{\otimes n}}{
	{}^{\otimes n}\bra\vac| 
	\mathbf{P}^+(\frak R_1) \mathbf{P}^-(\frak R_{-1})|\vac\ket^{\otimes n}\,.
	}.
\]
The insertion of permutation operators is seen as changing the state (the measure) over which we take the average\footnote{By cyclic permutation invariance of both the permutation twists and the $n$-copy vacuum state, we may put the permutation twists on the left or on the right, without changing the result. Therefore, the state still is real-valued on hermitician operators.}.

The first expected clustering property is that, when the points $\bm x_j$ are far from the boundaries of the regions $\p\frak R_1$ and $\p\frak R_{-1}$, we recover the vacuum state. That is,
for every $k\geq 1$ and every set $\{\bm p_1,\ldots,\bm p_k\}$ of elements of $\Lambda_d$,
there exists a function $U(\bm x_1,\ldots,\bm x_k)>0$ and a number $V>0$ such that for every set $\{\bm x_1,\ldots,\bm x_k\}$ of elements of $\frak M$, and for every regions $\frak R_1$ and $\frak R_{-1}$ as described above,
\beqa
	&&\Big|{}_{\mathbf P}\bra
	\Or_{\bm p_1}^{(j_1)}(\bm x_1)
	\cdots
	\Or_{\bm p_k}^{(j_k)}(\bm x_k)
	\ket_{\mathbf P}  - 
	{}^{\otimes n}\bra\vac| 
	\Or_{\bm p_1}^{(j_1)}(\bm x_1)
	\cdots
	\Or_{\bm p_k}^{(j_k)}(\bm x_k)|\vac\ket^{\otimes n}
	\Big| \n &&
	< \ U(\bm x_1,\ldots,\bm x_k)\,\exp\big[-V
	{\rm dist}(\{\bm x_i:i\in I_k\},\p\frak R_1\cup\p\frak R_{-1})\big]\,.
	\label{ass}
\eeqa
This clustering at large distance between the points $\bm x_i$ and the boundary of the regions $\mathfrak R_1$ and $\mathfrak R_{-1}$ indicates that the operator $\mathbf{P}^+(\frak R_1) \mathbf{P}^-(\frak R_{-1})$ is essentially supported on $\p\frak R_1\cup\p\frak R_{-1}$. This is expected, as the cyclic permutation of copies is a symmetry of the $n$-copy QFT, and thus $\mathbf{P}^+(\frak R_1) \mathbf{P}^-(\frak R_{-1})$ is a twist operator supported on the boundary of the permutation region.

Second, the function $U(\bm x_1,\ldots,\bm x_k)$ can also be bounded. This is because we expect the new state ${}_{\mathbf P}\bra\cdots \ket_{\mathbf P}$, like the vacuum, to still satisfy the clustering property of local fields. The initial observation is that, if the points $\bm x_i$ are all very far from each other, the normalised correlation function factorises into the averages, in this new state, of each local observable. By the $\Z_2$ symmetry of the Klein-Gordon theory, acting as $\Or_{\bm p}(\bm x)\mapsto -\Or_{\bm p}(\bm x)$ and preserving the vacuum and the permutation operators, the result vanishes. This vanishing is also exponential. We then conclude that the only way not to have a vanishing result is if for every $i\in I_k$ there exists a $j\in I_k$ such that ${\rm dist}(\bm x_i,\bm x_j)$ is finite, so that clustering occurs in groups of local fields with nonzero averages. In order to bound the resulting exponential decay due to these various groups, we can simply sum over every minimal distance between a point $\bm x_i$ and the rest. Hence there exists a $U_0>0$ and $W>0$ such that
\beq\label{boundU}
	U(\bm x_1,\ldots,\bm x_k) <
	U_0\exp\Big[-W
	\sum_{i\in I_k}{\rm dist}\big(\bm x_i,\{\bm x_j:j\in I_k\setminus\{i\}\}\big)
	\Big]\,,
\eeq
for all $\{\bm x_i\in\frak M\}$.

%
Using these properties, we show the following, which gives results for the replica logarithmic negativity increment for the entanglement between the scaled regions $\lambda \frak R_1$ and $\lambda \frak R_{-1}$, and the R\'enyi entanglement entropy increment for the entanglement  between the scaled regions $\lambda \frak R_1$ and $\lambda \frak M\setminus \lambda \frak R_1$, in the limit where the scaling $\lambda$ tends to infinity.
 
\begin{theorem} Let $\mathsf{p} = \{\bm p_1,\ldots,\bm p_k\}$ with $\bm p_j\in\R^d$, and denote by $[\bm p_j]_\lambda$ the point in $\lambda^{-1}\Lambda_d$ that is nearest to $\bm p_j$, and by $[\mathsf{p}]_\lambda = \{[\bm p_1]_\lambda,\ldots,\bm [\bm p_k]_\lambda\}$. Without loss of generality, assume that $\mathsf{p}$ is formed by groups of identical momenta, $\bm p_1=\ldots = \bm p_{k_1}$, $\bm p_{k_1+1}=\ldots = \bm p_{k_1+k_2}$, $\cdots$ with $\sum_i k_i = k$, and with momenta belonging to different groups being different.  Let $r_{\ell} = {\rm Vol}({\frak R}_\ell)/{\rm Vol}({\frak M})$ for $\ell\in\{1,-1\}$ and $r_0 = 1-r_1-r_{-1}$. Then
\beq\label{theoneg}
	\lim_{\lambda\to\infty}
	\Delta\mathcal{E}_n^{|[\mathsf p]_{\lambda}\ket}(\lambda{\frak R}_1,\lambda{\frak R}_{-1};\lambda{\frak M})
	= \sum_{i} \mathcal{E}_n^{|\Psi_{k_i}(r_1,r_0,r_{-1})\ket}\,.
\eeq
Let $r_1 = {\rm Vol}({\frak R}_1)/{\rm Vol}({\frak M})$ and $r_0 = 1-r_1$. Then
\beq\label{theoent}
	\lim_{\lambda\to\infty}
	\Delta S_n^{|[\mathsf p]_{\lambda}\ket}(\lambda{\frak R}_1;\lambda{\frak M})
	= \sum_{i} S_n^{|\Psi_{k_i}(r_1,r_0)\ket'}\,.
\eeq
\end{theorem}
Recall the expressions for the qubit entanglement quantities $\mathcal{E}_n^{|\Psi_{k_i}(r_1,r_0,r_{-1})\ket}$ and $S_n^{|\Psi_{k_i}(r_1,r_0)\ket'}$ in Theorems \ref{thformula} and \ref{thmain}.

\subsection{Proof}

We concentrate on the replica negativity; the entanglement entropy is obtained again as a special case. The idea of the proof is to use the expression \eqref{QFTE}, where the particle states are explicitly written in terms of local operators as per \eqref{Ap}. We then use the exchange relations \eqref{QFTexc} on the local operators, and evaluate the leading large-$\lambda$ behaviour using the clustering properties. There, the expression is transformed into a vacuum expectation value, with appropriately shifted copy indices, very similar to \eqref{qtytoe} obtained in the qubit analysis. The creation and annihilation operators are instead local fields, but the evaluation is again by Wick's theorem. We then analyse the Wick contractions, finding a structure similar to that obtained by evaluating \eqref{qtytoe}. Besides the re-writing into local fields, there is one additional subtlety, as particles carry momenta. We show that particles with different momenta, in the large-$\lambda$ limit, do not Wick contract, and thus the result factorises into groups of equal momenta. In each group, the ensuing graph analysis goes through essentially unchanged.

Using \eqref{Ap}, 
\beqa
	\lefteqn{{}^{\otimes n}\bra[\mathsf{p}]_{\lambda}| \mathbf{P}^+(\lambda \frak R_1) \mathbf{P}^-(\lambda \frak R_{-1})|[\mathsf{p}]_{\lambda}\ket^{\otimes n}}&&
	\n &=&
	\frc1{\lambda^{dkn}{\rm Vol}(\frak M)^{kn}}\,\lt(\prod_{j=1}^k \prod_{i=1}^n \int_{\lambda \frak M}
	\dd^d\bm x_j^{(i)}
	\int_{\lambda \frak M}
	\dd^d\bm y_j^{(i)}\rt)\,
	\re^{-\ri \sum_{j=1}^k \sum_{i=1}^n [\bm p_j]_{\lambda}\cdot (\bm x_j^{(i)}-\bm y_j^{(i)})}
	\,\times\n && \times \
	{}^{\otimes n}\bra\vac|
	\Bigg(\prod_{j=1}^k \prod_{i=1}^n
	\Or_{[\bm p_j]_\lambda}^{(i)}(\bm x_j^{(i)})\Bigg)
	\,
	\Bigg(\prod_{j'=1}^k \prod_{i'=1}^n
	\left[\Or_{[\bm p_{j'}]_{\lambda}}^{(i'+\chi(y_{j'}^{(i')}\in\lambda \frak R_1) -
	\chi(y_{j'}^{(i')}\in\lambda \frak R_{-1}))}(\bm y_{j'}^{(i')})\right]^\dag\Bigg)
	\,\times\n &&\quad\qquad \times \
	\mathbf{P}^+(\lambda \frak R_1) \mathbf{P}^-(\lambda \frak R_{-1})|\vac\ket^{\otimes n}\,.\no
\eeqa
The first step is to show that, in the large $\lambda$ limit, the expectation in the integrand can be factorised into a vacuum expectation value of the local observables, times that of the permutation twists. This is done by using \eqref{ass}, and arguing that the integration over the bulk of the regions, far from the boundaries $\p\frak R_{1,-1}$, is that which dominates.
We therefore consider the difference
\beqa
	\lefteqn{D_\lambda(\{\bm x^{(i)}_j,\bm y^{(i')}_{j'}\}) =} && \n
	&& 
{}^{\otimes n}\bra\vac|
	\Bigg(\prod_{j=1}^k \prod_{i=1}^n
	\Or_{[\bm p_j]_\lambda}^{(i)}(\bm x_j^{(i)})\Bigg)
	\,
	\Bigg(\prod_{j'=1}^k \prod_{i'=1}^n
	\left[\Or_{[\bm p_{j'}]_{\lambda}}^{(i'+\chi(y_{j'}^{(i')}\in\lambda \frak R_1) -
	\chi(y_{j'}^{(i')}\in\lambda \frak R_{-1}))}(\bm y_{j'}^{(i')})\right]^\dag\Bigg)
	\,\times\n && \quad\qquad \times \
	\mathbf{P}^+(\lambda \frak R_1) \mathbf{P}^-(\lambda \frak R_{-1})|\vac\ket^{\otimes n}
	\; / \;
	{}^{\otimes n}\bra\vac|
	\mathbf{P}^+(\lambda \frak R_1) \mathbf{P}^-(\lambda \frak R_{-1})|\vac\ket^{\otimes n}
	\n && -\  
	{}^{\otimes n}\bra\vac|
	\Bigg(\prod_{j=1}^k \prod_{i=1}^n
	\Or_{[\bm p_j]_\lambda}^{(i)}(\bm x_j^{(i)})\Bigg)
	\,
	\Bigg(\prod_{j'=1}^k \prod_{i'=1}^n
	\left[\Or_{[\bm p_{j'}]_{\lambda}}^{(i'+\chi(y_{j'}^{(i')}\in\lambda \frak R_1) -
	\chi(y_{j'}^{(i')}\in\lambda \frak R_{-1}))}(\bm y_{j'}^{(i')})\right]^\dag\Bigg)|\vac\ket^{\otimes n}\no\,,
\eeqa
and, recalling \eqref{ass}, we have
\beqa
	\lefteqn{
		\lt|\frc1{\lambda^{dkn}{\rm Vol}(\frak M)^{kn}}\,\lt(\prod_{j=1}^k \prod_{i=1}^n \int_{\lambda \frak M}
	\dd^d\bm x_j^{(i)}
	\int_{\lambda \frak M}
	\dd^d\bm y_j^{(i)}\rt)\,
	\re^{-\ri \sum_{j=1}^k \sum_{i=1}^n [\bm p_j]_{\lambda}\cdot (\bm x_j^{(i)}-\bm y_j^{(i)})}
D_\lambda(\{\bm x^{(i)}_j,\bm y^{(i')}_{j'}\})\rt|}
	\n &<&
		\frc1{\lambda^{dkn}{\rm Vol}(\frak M)^{kn}}\,\lt(\prod_{j=1}^k \prod_{i=1}^n \int_{\lambda \frak M}
	\dd^d\bm x_j^{(i)}
	\int_{\lambda \frak M}
	\dd^d\bm y_j^{(i)}\rt)\,
\lt|D_\lambda(\{\bm x^{(i)}_j,\bm y^{(i')}_{j'}\})\rt| \n &<&
 E_\lambda   \quad =\quad \frc1{\lambda^{dkn}{\rm Vol}(\frak M)^{kn}}\,\prod_{j=1}^k \prod_{i=1}^n \lt(\int_{\lambda \frak M}
	\dd^d\bm x_j^{(i)}
	\int_{\lambda \frak M}
	\dd^d\bm y_j^{(i)}\rt)
	\,\times \n && \qquad\times \
	\prod_{j=1}^k \prod_{i=1}^n\Big(
	U(\{\bm x_j^{(i)},\bm y_j^{(i)}\})\,\exp\big[-V
	{\rm dist}(\{\bm x_j^{(i)},\bm y_j^{(i)}:j\in I_k,\,i\in I_n\},\p\frak R_1\cup\p\frak R_{-1})
	\big]\Big)\,,\n
	\label{corrterm}
\eeqa
where the function $U$ is bounded as per \eqref{boundU}. We now show that
\beq
	\lim_{\lambda\to\infty}E_\lambda = 0\,.
\eeq
Thanks to the bound \eqref{boundU}, the integrals over the variables $\bm x_j^{(i)}$'s and $\bm y_j^{(i)}$'s, lying on the manifold $(\lambda \frak M)^{\times 2kn}$ of dimension $2dkn$, are supported, with exponential accuracy, over a submanifold which is at most of dimension $dkn$. Indeed, since every variable must lie near to at least one other, one forms pairs or larger groups of nearby variables; forming pairs leads to the largest submanifold, and the dimension of this submanifold is that of the original integration manifold $(\lambda \frak M)^{\times 2kn}$ divided by 2. Further, thanks to the exponential in \eqref{corrterm}, one further restricts all variables to lie near the boundary of $\frak R_1$ or $\frak R_{-1}$, thus near a submanifold of codimension 1. The remaining integration region is therefore an effectively finite neighborhood (thanks to exponential accuracy) of a submanifold of $(\lambda \frak M)^{\times 2kn}$ of dimension $dkn-1$. As $\lambda\to\infty$, this scales like $\lambda^{dkn-1}$. Because of the factor $\lambda^{dkn}$ in the denominator in \eqref{corrterm}, the result vanishes as $\lambda\to\infty$.

Therefore, using \eqref{ass}, we find
\beqa
	\lefteqn{\frc{{}^{\otimes n}\bra[\mathsf{p}]_{\lambda}| \mathbf{P}^+(\lambda \frak R_1) \mathbf{P}^-(\lambda \frak R_{-1})|[\mathsf{p}]_{\lambda}\ket^{\otimes n}}{
	{}^{\otimes n}\bra\vac|\mathbf{P}^+(\lambda \frak R_1) \mathbf{P}^-(\lambda \frak R_{-1})|\vac\ket^{\otimes n}
	}}&&
	\n &=&
	\frc1{\lambda^{dkn}{\rm Vol}(\frak M)^{kn}}\,\lt(\prod_{j=1}^k \prod_{i=1}^n \int_{\lambda \frak M}
	\dd^d\bm x_j^{(i)}
	\int_{\lambda \frak M}
	\dd^d\bm y_j^{(i)}\rt)\,
	\re^{-\ri \sum_{j=1}^k \sum_{i=1}^n [\bm p_j]_{\lambda}\cdot (\bm x_j^{(i)}-\bm y_j^{(i)})}
	\,\times\n && \times \
	{}^{\otimes n}\bra\vac|
	\Bigg(\prod_{j=1}^k \prod_{i=1}^n
	\Or_{[\bm p_j]_\lambda}^{(i)}(\bm x_j^{(i)})\Bigg)
	\,
	\Bigg(\prod_{j'=1}^k \prod_{i'=1}^n
	\left[\Or_{[\bm p_{j'}]_{\lambda}}^{(i'+\chi(y_{j'}^{(i')}\in\lambda \frak R_1) -
	\chi(y_{j'}^{(i')}\in\lambda \frak R_{-1}))}(\bm y_{j'}^{(i')})\right]^\dag\Bigg)|\vac\ket^{\otimes n}
	\n && 
	+\; O(\lambda^{-1})\,.
	\label{resuy}
\eeqa
The correlation function in \eqref{resuy} is evaluted by Wick's theorem. Every Wick contraction between operators $\Or_{[\bm p_j]_\lambda}^{(i)}(\bm x_j^{(i)})$ and $\Or_{[\bm p_{\t j}]_\lambda}(\bm x_{\t j}^{(\t i)})$ gives exactly, under integrations over $\bm x_j^{(i)}$ and $\bm x_{\t j}^{(\t i)}$, the overlap $\bra\vac|A_{p_j}A_{p_{\t j}}|\vac\ket$, which vanishes. Hence all operators  $\Or_{[\bm p_j]_\lambda}^{(i)}(\bm x_j^{(i)})$ must be contracted with operators $\left[\Or_{[\bm p_{j'}]_\lambda}^{(i'+\chi(y_{j'}^{(i')}\in\lambda \frak R_1) -
	\chi(y_{j'}^{(i')}\in\lambda \frak R_{-1}))}(\bm y_{j'}^{(i')})\right]^\dag$.

These contractions may be evaluated as follows. We note that
\beqa
	\delta_{i,i'}\delta_{[\bm p]_\lambda,[\bm p']_\lambda}
	&=& {}^{\otimes n}\bra\vac| A_{ [\bm p]_\lambda}^{(i)}
	\big(A_{[\bm p']_\lambda}^{(i')}\big)^\dag
	|\vac\ket^{\otimes n} \n
	&=&
	\frc1{\lambda^d {\rm Vol}(\frak M)}
	\int_{\lambda\frak M} \dd^d \bm x 
	\int_{\lambda\frak M} \dd^d \bm y\,
	\re^{-\ri [\bm p]_\lambda \cdot
	\bm x+\ri [\bm p']_\lambda \cdot
	\bm y}\;
	{}^{\otimes n}\bra\vac|
	\Or^{(i)}_{[\bm p]_\lambda}(\bm x)
	\left[\Or^{(i')}_{[\bm p']_\lambda}(\bm y)\right]^\dag
	|\vac\ket^{\otimes n}\,.\no
\eeqa
Further, as per the discussion above \eqref{boundU}, by standard results in the massive free boson, the function ${}^{\otimes n}\bra\vac|\Or^{(i)}_{[\bm p]}(\bm x) \left[\Or^{(i')}_{[\bm p']}\right]^\dag(\bm y) |\vac\ket^{\otimes n}$ is exponentially decaying with ${\rm dist}(\bm x,\bm y)$, and, by translation invariance of the vacuum and factorisation into the $n$ copies, it is a function of $\bm x - \bm y$ only and vanishes if $i\neq i'$. In particular, we obtain
\beq\label{ooo}
	\int_{\lambda\frak M}\dd^d \bm x \,
	\re^{-\ri [\bm p]_\lambda\cdot\bm x}
	\ {}^{\otimes n}\bra\vac|
	\Or^{(i)}_{[\bm p]_\lambda}(\bm x)
	\left[\Or^{(i)}_{[\bm p]_\lambda}(\bm 0)\right]^\dag
	|\vac\ket^{\otimes n}
	= 1\,.
\eeq

Because of diagonality in the space of copy indices, in any given Wick contraction between  $\Or_{[\bm p_j]_\lambda}^{(i)}(\bm x_j^{(i)})$ and $\left[\Or_{[\bm p_{j'}]_\lambda}^{(i'+\chi(y_{j'}^{(i')}\in\lambda \frak R_1) -
	\chi(y_{j'}^{(i')}\in\lambda \frak R_{-1}))}(\bm y_{j'}^{(i')})\right]^\dag$ occurring in \eqref{resuy}, for any fixed $i,i',j,j'$ giving rise to a nonzero Wick contraction,  the region of integration of the $\bm y_{j'}^{(i')}$ coordinate is restricted to a region within $\frak M$, as per the condition of equality of copy numbers,
\[
i = i'+\chi(y_{j'}^{(i')}\in\lambda \frak R_1) -
	\chi(y_{j'}^{(i')}\in\lambda \frak R_{-1}).
\]
Let us therefore consider one such integrated contraction, say with $\bm y$ restricted to some region $\lambda \frak R$:
\beq
	C_{\bm p,\bm p'}^{i,i'}(\frak R) = \lim_{\lambda\to\infty}
	\frc1{\lambda^d {\rm Vol}(\frak M)}
	\int_{\lambda\frak M} \dd^d \bm x 
	\int_{\lambda\frak R} \dd^d \bm y\,
	\re^{-\ri [\bm p]_\lambda \cdot
	\bm x+\ri [\bm p']_\lambda \cdot
	\bm y}\;
	{}^{\otimes n}\bra\vac|
	\Or^{(i)}_{[\bm p]_\lambda}(\bm x)
	\left[\Or^{(i')}_{[\bm p']_\lambda}(\bm y)\right]^\dag
	|\vac\ket^{\otimes n}\,.
\eeq
By the properties of the two-point function mentioned above, this equals
\beq
	C_{\bm p,\bm p'}^{i,i'}(\frak R) = \lim_{\lambda\to\infty}
	G_\lambda
	\int_{\lambda\frak M} \dd^d \bm x \,
	\re^{-\ri [\bm p]_\lambda \cdot
	\bm x}\ 
	{}^{\otimes n}\bra\vac|
	\Or^{(i)}_{[\bm p]_\lambda}(\bm x)
	\left[\Or^{(i')}_{[\bm p']_\lambda}(\bm 0)\right]^\dag
	|\vac\ket^{\otimes n}\,,
\eeq
where
\beq\label{Glambda}
	G_\lambda = \frc1{\lambda^d {\rm Vol}(\frak M)}
\int_{\lambda\frak R} \dd^d \bm y\,
	\re^{\ri ([\bm p']_\lambda- [\bm p]_\lambda) \cdot
	\bm y}\,.
\eeq

We now show that 
\beq
	G_\lambda = \lt\{\ba{ll}\displaystyle
	\frc{{\rm Vol}(\frak R)}{{\rm Vol}(\frak M)} &
	(\bm p=\bm p')\\ 0 & (\mbox{otherwise})\,.
	\ea\rt.
\eeq
The idea is that in the integral \eqref{Glambda}, if the momenta are different, then the integrand is oscillatory, and it integrates to zero on every complete period. As $\lambda\to\infty$, the period stays finite while the region grows. The integral is zero within the bulk of the region $\lambda\mathfrak R$, and it receives nonzero contributions only on an integration region near the boundary of $\lambda\mathfrak R$, where the integration is not on a full period (the period being broken by the region's boundary).

A precise proof is as follows. If $\bm p\neq \bm p'$, then for all $\lambda$ large enough, $[\bm p]_\lambda\neq[\bm p']_\lambda$. Consider $\lambda$ large enough, and one direction $j\in\{1,\ldots,d\}$ where there is a difference: $[p_j]_\lambda\neq[p_j']_\lambda$. Let us divide the region $\lambda \frak M$, in this direction, into slices $\alpha$ (which extend in all directions $j'\neq j$) of width $\delta y_j = 2\pi |[p_j]_\lambda-[p'_j]_\lambda|^{-1}$; this is the period of the oscillatory exponential in this particular direction. The slices are the subsets $[0,\lambda L_1]\times \cdots\times [\alpha\delta y_j , (\alpha+1)\delta y_j]\times \cdots\times [0,\lambda L_d]\in \lambda\frak M$ and $\alpha$ is in a subset of $\Z$ such that these cover $\frak M$ (note that $\frak M$ is built out of an integer number of complete slices). On every  slice $\alpha$, at every point along it where the width is fully contained within $\lambda \frak R$, that is $(y_1,\ldots,[\alpha\delta y_j , (\alpha+1)\delta y_j],\ldots, y_d)\subset\lambda\frak R$, the contribution to the above integral vanishes by integration over $y_j$. As $\lambda\to\infty$, the set of all such segments $(y_1,\ldots,[\alpha\delta y_j , (\alpha+1)\delta y_j],\ldots, y_d)\subset\lambda\frak R$ covers all of $\frak R$ excepts for a neighbourhood of width at most $\delta y_j$ of its boundary $\p\frak R$. That is, if $\bm p\neq\bm p'$, for $\lambda$ large enough, we can bound $G_\lambda$ as
\beq
	|G_\lambda| < \frc{2\pi}{|[p_j]_\lambda-[p'_j]_\lambda|}
	\frc1{\lambda^d {\rm Vol}(\frak M)}
\int_{\lambda\p\frak R} \dd^{d-1} \bm y
	\stackrel{\lambda\to\infty} = 0\,.
\eeq
On the other hand, clearly, for $\bm p=\bm p'$, we have
\beq
	G_\lambda = \frc{{\rm Vol}(\frak R)}{{\rm Vol}(\frak M)}.
\eeq
As a consequence, using \eqref{ooo}, we obtain
\beq\label{contract}
	C_{\bm p,\bm p'}^{i,i'}(\frak R) = \delta_{i,i'}
	\delta_{\bm p,\bm p'}
	\frc{{\rm Vol}(\frak R)}{{\rm Vol}(\frak M)}\,.
\eeq

From this point on, using the Wick contraction \eqref{contract}, the discussion following \eqref{qtytoe}  goes through, up to two differences: (1) the extra condition that momenta in a contraction must take the same value, and (2) the contribution $\frc{{\rm Vol}(\frak R_\ell)}{{\rm Vol}(\frak M)}$ for every edge, instead of $L_\ell$. The requirement that momenta must agree gives a product, over different groups of equal momenta, of the result obtained there, in terms of graph partition functions:
\beq
	\lim_{\lambda\to\infty}
	\exp\big[\Delta\mathcal{E}_n^{|[\mathsf p]_{\lambda}\ket}(\lambda{\frak R}_1,\lambda{\frak R}_{-1};\lambda{\frak M})\big]
	= \prod_{i} \frc{p_{k_i,n}(r_1,r_0,r_{-1})}{(k_i!)^n}\,.
\eeq
From Theorem \ref{thmain} the result  \eqref{theoneg} follows. Setting $\frak R_{-1}= \emptyset$, \eqref{theoent} follows similarly.

\eproof

\section{Conclusion}\label{sectconclu}

We have established exact relations between the replica logarithmic negativity and R\'enyi entanglement entropy of certain qubit states representing uniform distribution of particles, and certain graph partition functions. The vertices and edges of the graphs have a natural interpretation in terms of the connectivity of the manifold, that naturally emerges in QFT, associated to the permutation-twist representation of entanglement measures. The result is however general, and applies to qubit states without the need for a QFT.

We have also evaluated the increment of replica logarithmic negativity and R\'enyi entropy in many-particle states with respect to the vacuum, in free bosonic QFT of any dimension on the hypertorus, in the limit where the volumes of the hypertorus and of the regions are large. The result is exactly that found in \cite{the4ofus3} in the one-dimensional case (and proposed there to be more general), equating these to the same quantities for the qubit states we discussed. The present paper thus gives a full proof of the general result in free bosonic QFT, showing that it holds independently of the connectivity, dimensionality and shape of the regions.

The proof involves different principles from those used in \cite{the4ofus3}, where the results were based on form factors of twist fields. Instead, here we use clustering properties of local fields along with the fundamental exchange relations characterising permutation twists in a given twist sector.

A number of generalisations should be immediate.
First, the QFT proof is based on very general properties. These properties are expected to hold in  states other than the vacuum, and in other free models. It is a simple matter to generalise, for instance, to cases with many particle types. More interestingly, we expect similar results to hold in the so-called generalised Gibbs ensembles \cite{Essler2016}, with density matrix of the form $\exp\big[-\sum_{\bm p\in\Lambda_d} w(\bm p) A_{\bm p}^\dag A_{\bm p}\big]$ for appropriate $w(\bm p)$. There, ``particle" and ``hole" excitations can be defined naturally as action of creation and annihilation operators \cite{DfiniteTff2005,DfiniteTff2007,ChenDoyon2014}, essentially via the Gelfand-Naimark-Segal mechanism where the space of operators is seen as the Hilbert space of the theory and the vacuum represents the original mixed state itself. This is relevant, as the negativity is a good measure of entanglement in mixed states such as GGEs. 

Second, it does not seem to be essential to take the fundamental operator in the twist sector in order to get the result: ``generalised" types of replica logarithmic negativities and R\'enyi  entropies, defined via descendants of such operators in the same twist sector, might lead to the same increment results. They would be related to more general partition functions in multi-sheeted manifolds, with insertion of fields at the boundaries of the regions $\frak R_1$, $\frak R_{-1}$.

Third, a generalisation of the results and proofs to integrable models appears to be possible as well. Indeed, in integrable models, because of the presence of stable quasi-particles, one can construct fields  which create asymptotic states and which, although not local, have strong enough quasi-locality properties. One could then use such fields, along with clustering properties, in order to adapt the proof we presented here. One might also hope that the same works in certain non-integrable models, below the particle-creation threshold or if there is particle conservation.

Finally, it is also a simple matter to generalise to any other product of permutation twists associated to more complicated connectivities. In all cases, the uniform-particle qubit states will lead to graph  partition functions  \eqref{partitionfunctiongeneral}, and the QFT increments, in the large volume limit, will again be equal to the qubit-state results.

\paragraph{Acknowledgments:} Olalla A. Castro-Alvaredo, Benjamin Doyon, and Istv\'an M. Sz\'ecs\'enyi are grateful to EPSRC for funding through the standard proposal {\it ``Entanglement Measures, Twist Fields, and Partition Functions in Quantum Field Theory"} under reference numbers EP/P006108/1 and EP/P006132/1. Cecilia De Fazio gratefully acknowledges funding from the School of Mathematics, Computer Science and Engineering of City, University of London through a PhD Studentship. This research was supported in part by Perimeter Institute for Theoretical Physics. Research at Perimeter Institute is supported by the Government of Canada through the Department of Innovation, Science and Economic Development and by the Province of Ontario through the Ministry of Research, Innovation and Science.

\appendix
\section{Graphs, partitions and negativity: examples}\label{appgraphs}
In this section we discuss in details two examples of the formula (\ref{formula}).

\subsection{The case $k=1$: a single particle excitation}
For $k=1$ we have that 
\beq
\mathcal{E}_n^{|\Psi_1(r_1,r_0,r_{-1})\ket}=\log\left(\sum_{p=-1}^1\sum_{\sigma=\max(0,-n p)}^{[\frac{n(1-p)}{2}]} A_{p,\sigma} r_1^{np+\sigma} r_0^{n(1-p)-2\sigma} r_{-1}^\sigma\right)\,,
\label{sum1}
\eeq
where the sums have now been restricted only to non-vanishing contributions, and
\beq
A_{p,\sigma}=\sum_{\{k_1,\ldots,k_n\}\in P_n(\sigma)} \prod_{j=1}^n \frac{1}{(p+k_j)!(1-p-k_{j+1}-k_j)! k_{j+1}!}\,,
\label{pro1}
\eeq
where here and below, $k_{n+1} \equiv k_1$. Therefore there are only three possible values of $p$ to consider. 
\begin{enumerate}
\item 
If $p=1$ then the only way the product in (\ref{pro1}) can be non-vanishing is if all the $k_j=0$ (otherwise, the middle factorial will involve a negative value for some $j$). In this case $\sigma=0$ and $A_{0,0}=1$ and this gives the contribution $r_1^{n}$.
\item If $p=0$ then $1-p-k_{j+1}-k_j\geq 0$ if and only if the partition consists entirely of 0s and 1s, with the 1s being non-consecutive. This can only be achieved if $0\leq \sigma \leq \frac{n}{2}$. We then have that $A_{0,\sigma}$ is exactly the number of partitions of $\sigma$ into $n$ parts, all of which are either 0 or 1 and where there are no consecutive 1s. It is easy to show this number is precisely
\beq
A_{0,\sigma} :=Q_\sigma=\frac{n}{n-\sigma}  \left(\begin{array}{c}
n-\sigma\\
\sigma
\end{array}\right),
\label{ans}
\eeq
and the sum over $\sigma$ in (\ref{sum1}) for $p=0$ then becomes
\beq
\sum_{\sigma=0}^{[\frac{n}{2}]} Q_{\sigma}r_0^{n-2\sigma} r_1^\sigma r_{-1}^\sigma\,.
\eeq
Above we introduced the notation $Q_\sigma$ because this coefficient will feature several times from now on. 
\item If $p=-1$ then the presence of the factorial $(-1+k_j)!$ requires that  $k_j\geq 1$ for all $j$. The presence of the factorial $(2-k_{j+1}-k_j)!$ restricts this condition to simply $k_j=1$. From this condition is follows that $\sigma=n$ and that, in this case, only the partition $\{1,1,\ldots,1\}$ contributes. The corresponding coefficient is $A_{-1,n}=1$ and this gives the contribution  $r_{-1}^n$.
\end{enumerate}
Therefore, for $k=1$ we can write the replica logarithmic negativity as
\beq
\mathcal{E}_n^{|\Psi_1(r_1,r_0,r_{-1})\ket}=\log\left(r_1^n+r_{-1}^n + \sum_{\sigma=0}^{[\frac{n}{2}]} A_{0,\sigma} r_0^{n-2\sigma} r_1^\sigma r_{-1}^\sigma \right)\,,
\label{fork1}
\eeq
which is the formula reported in \cite{the4ofus3}, where it was derived from a form factor calculation and also from the explicit diagonalization of the partially transposed reduced density matrix. 

\begin{figure}[h!]
 \begin{center} 
 \includegraphics[width=9cm]{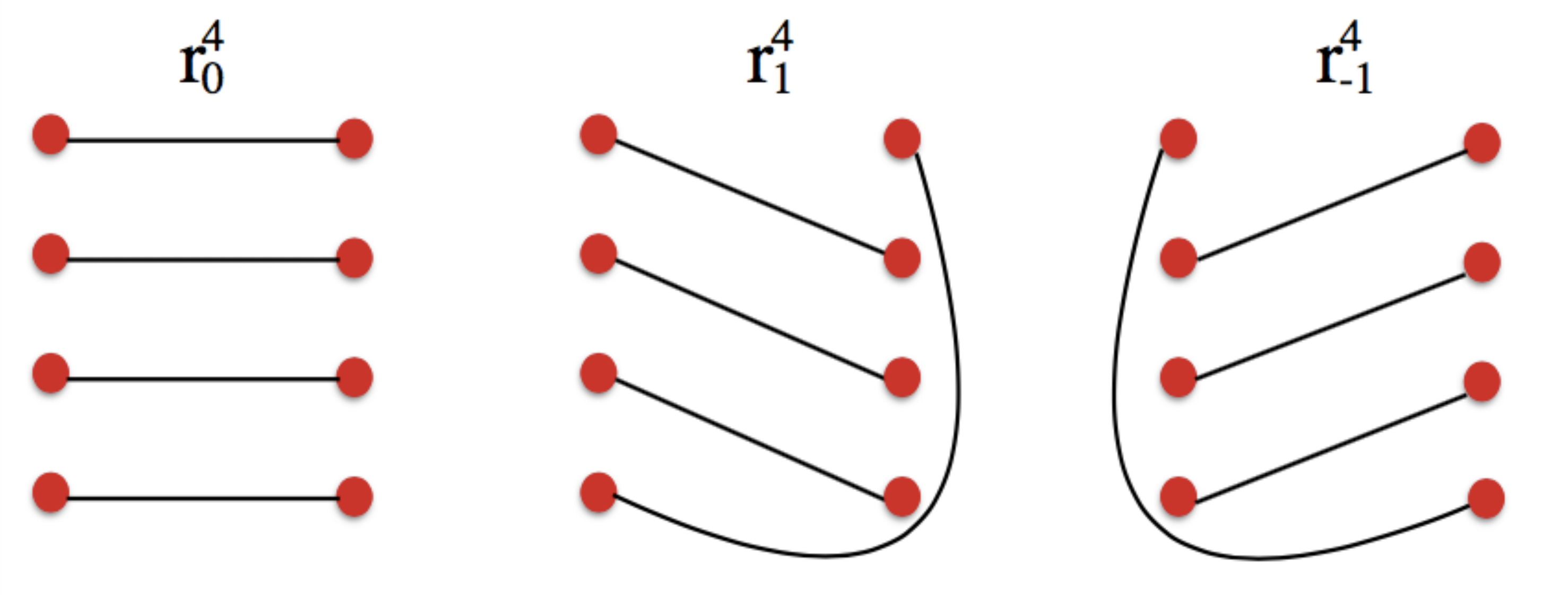} 
  \includegraphics[width=12cm]{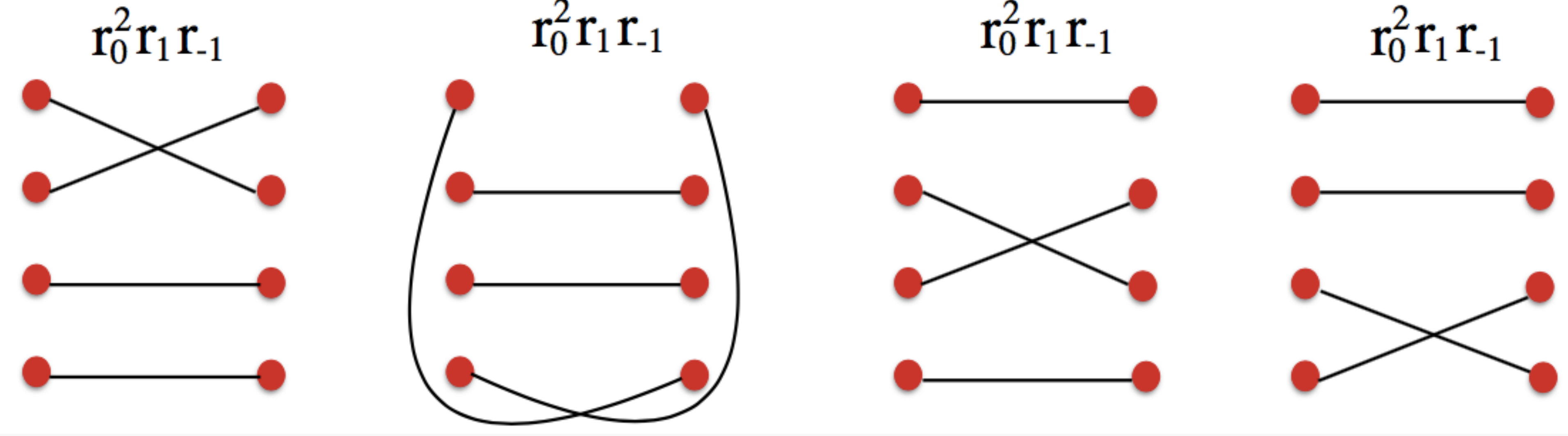} 
   \includegraphics[width=6cm]{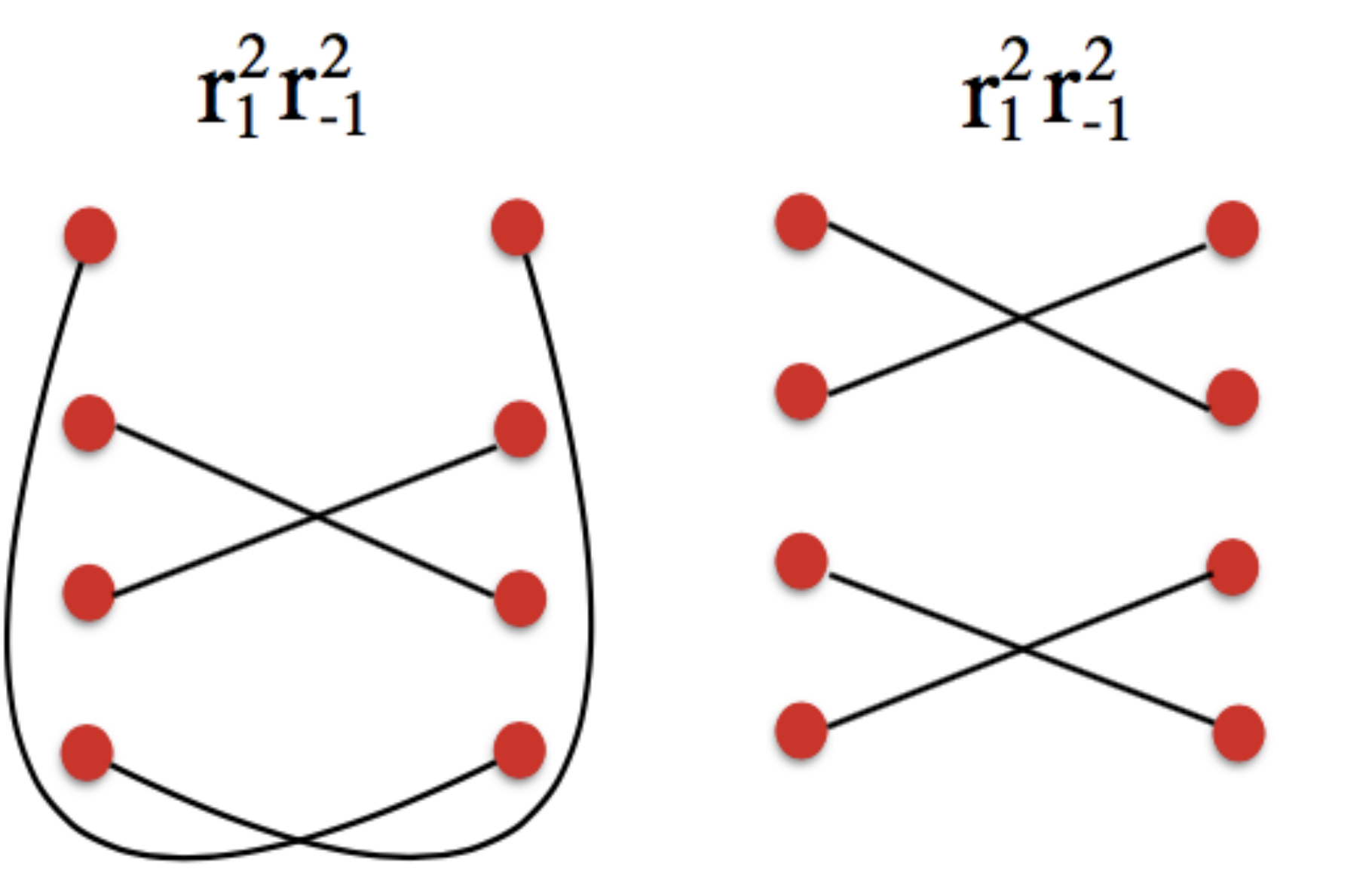} 
  \end{center} 
 \caption{All contributing graphs for $k=1$, $n=4$. In this case the simple addition of individual contributions reproduces the exponential of the function (\ref{fork1n4}).}
 \vspace{1cm}
 \label{negative12} 
 \end{figure}

The various terms in (\ref{fork1}) admit also a graphical representation, where different elements of the graphs are assigned weights $r_1$, $r_0$, or $r_{-1}$ as shown in Fig.~\ref{negative1}. 
Let us consider as an example, the case $k=1, n=4$. In this case 
\beq
\mathcal{E}_4^{|\Psi_1(r_1,r_0,r_{-1})\ket}=\log\left(r_1^4+r_{-1}^4 + r_0^4+ 4 r_0^2 r_1 r_{-1}+ 2 r_1^2 r_{-1}^2 \right)\,,
\label{fork1n4}
\eeq
The terms in this formula are generated from the graphs in Fig.\ref{negative12}.

\subsection{The case $k=2$: excited state of two identical particles}
For $k=2$ we have the formulae:
\beq
\mathcal{E}_n^{|\Psi_2(r_1,r_0,r_{-1})\ket}=\log\left(\sum_{p=-2}^2\sum_{\sigma=\max(0,-n p)}^{[\frac{n(2-p)}{2}]} A_{p,\sigma} r_1^{np+\sigma} r_0^{n(2-p)-2\sigma} r_{-1}^\sigma\right)\,,
\label{sum2}
\eeq
where the sums have now been restricted only to non-vanishing contributions, and
\beq
A_{p,\sigma}=\sum_{\{k_1,\ldots,k_n\}\in P_n(\sigma)} \prod_{j=1}^n \frac{2}{(p+k_j)!(2-p-k_{j+1}-k_j)! k_{j+1}!}\,,
\label{pro2}
\eeq
therefore there are five possible values of $p$ to consider. 
\begin{enumerate}
\item  If $p=2$ the only partition of $\sigma$ that gives a non-vanishing coefficient is the partition where $k_j=0$ for all $j$. That is $\sigma=0$ and $A_{2,0}=1$ which gives the term $r_1^{2n}$.
\item If $p=1$ then the only partitions of $\sigma$ that give non-vanishing coefficients are those whose parts are either 0 or 1 and where there are no consecutive 1s. The latter condition means that $0\leq \sigma \leq \frac{n}{2}$ and 
\beq 
A_{1,\sigma}=2^{n-\sigma} \times (\mathrm{number \, of \, such \, partitions})=2^{n-\sigma}Q_\sigma\,,
\eeq
where $Q_\sigma$ was defined in (\ref{ans}). This gives the terms
\beq
\sum_{\sigma=0}^{[\frac{n}{2}]}2^{n-\sigma}Q_\sigma r_1^{n+\sigma} r_0^{n-2\sigma} r_{-1}^\sigma\,.
\label{adisum}
\eeq
\item If $p=0$ the only partitions that give non-vanishing contributions are those that have parts which are either 0, 1 or 2 and which have no consecutive 2s and no consecutive 1s and 2s. There are many such partitions and challenge is to count them all and work out their contributions according to (\ref{pro2}). 

The two simplest cases are the partition where all terms are 0s and the partition where all terms are 1s. In the first case $\sigma=0$ and $A_{0,0}=1$. This gives the contribution $r_0^{2n}$. In the second case we have that $\sigma=n$ and $A_{0,n}=2^n$ and this gives the term $2^n (r_1 r_{-1})^n$. An additional partition corresponding to $\sigma=n$ is obtained when $n$ is even and every other term is either a 0 or a 2. There are two partitions of this type and when present they will give and additional contribution so that $A_{0,n}=2^n+2$, in agreement with the results we already knew from \cite{the4ofus3}. 

More generally we can now consider all partitions including at least one 0 and 1s and 2s according to the constraints above. These correspond to $1\leq \sigma \leq n-1$.

 For $\sigma=1$ we have a single 1 and there are $n$ such partitions. The coefficient $A_{0,1}=\frac{2^n}{2^{n-2}} \times n$ and this gives the contribution $4n r_0^{2n-2} r_1 r_{-1}$.

For $\sigma=2$ we will now have partitions that contain either two 1s or a single 2 (with everything else being 0). There are $n$ partitions that contain a single 2 and they give a contribution to the coefficient $A_{0,2}$ which is given by $\frac{2^n}{2 \times 2^{n-1}} \times n$. The partitions that contain two 1s can be divided into those where the 1s are consecutive and those where they are not. There are $n$ partitions that contain two consecutive 1s and their contribution to $A_{0,2}$ is $\frac{2^n}{2^{n-3}} \times n$. Finally the number of partitions that contain two non-consecutive ones is given by the coefficient $ Q_2$ defined earlier and these give a contribution to $A_{0,2}$ which is given by $\frac{2^n}{2^{n-4}} \times Q_2$. So, overall
\beq
A_{0,2}= \frac{2^n}{2^{n-4}} \times Q_2+ \frac{2^n}{2^{n-3}} \times n + \frac{2^n}{2 \times 2^{n-1}} \times n = 8 n (n-3) + 8n+n=n(8n-15)\,,
\eeq
which gives the contribution $n(8n-15) r_0^{2n-4} (r_1 r_{-1})^2$.

For $\sigma=3$ we can have partitions that contain either three 1s or a 1 and a 2 that are not consecutive.  The partitions that contain a 1 and a 2 that are not consecutive contribute 
\beq
\frac{2^n} {2\times 2^{n-4} \times 2}\times (\mathrm{number \, of \, such \, partitions})= 4 \times(2 Q_2)=4n(n-3).
\eeq
The partitions consisting of three 1s need to be divided into those where there are no consecutive 1s, those where two 1s are consecutive and those where the three 1s are consecutive. 
The partitions that have no consecutive 1s give a contribution
\beq
\frac{2^n} {2^{n-6}}\times (\mathrm{number \, of \, such \, partitions})= 64 Q_3= \frac{32n(n-4)(n-5)}{3}.
\eeq
There are $n$ partitions where all three 1s are consecutive and they give a contribution
\beq
\frac{2^n}{2^{n-4}}\times n =16n.
\eeq
Finally, the number of partitions where two 1s are consecutive and one is not is given by the product of $n$ ways of placing two consecutive 1s times $n-4$ ways of placing the remaining 1. This gives a contribution
\beq
\frac{2^n}{2^{n-5}} \times {n(n-4)}={32n(n-4)}.
\eeq
So, the overall coefficient is
\beq
A_{0,3}=4n(n-3)+\frac{32n(n-4)(n-5)}{3}+16n+{32n(n-4)}=\frac{4n}{3}(8n^2-45n+67).
\eeq
This gives the contribution $\frac{4n}{3}(8n^2-45n+67) r_0^{2n-6} (r_1 r_{-1})^3$.

One can proceed similarly to higher values of $\sigma$ and obtain increasingly complicated formulae for the coefficients as reported in Appendix B of \cite{the4ofus3}. However, there is no obvious pattern in $n$ emerging. Interestingly all coefficients $A_{0,\sigma}$ return integer values, even though this is also not obvious from the formulae.
\item If $p=-1$ we need partitions where all parts are at least 1, so the smallest allowed value of $\sigma$ is $\sigma=n$ corresponding to the partition into 1s. This gives $A_{-1,n}=2^n$ and corresponds to the term $2^n(r_{-1} r_0)^n$. This is the ``minimal" partition but there will be others. In fact, it is easy to argue that additional contributions will give rise to a very similar sum as (\ref{adisum}) with the roles of $r_1$ and $r_{-1}$ exchanged:
\beq
\sum_{\sigma=0}^{[\frac{n}{2}]}2^{n-\sigma}Q_\sigma r_{-1}^{n+\sigma} r_0^{n-2\sigma} r_{1}^\sigma\,,
\eeq
\item If $p=-2$ we then need all parts of any contributing partition to be 2 or larger, for the first factorial in the denominator of (\ref{pro2}) to be finite. All parts must be less or equal 2 for the second factorial to remain positive. This means only the partition consisting entirely of 2s will contribute. This corresponds to $\sigma=2n$ and $A_{-2,2n}=1$ and gives the contribution $r_{-1}^{2n}$. 
\end{enumerate}
 \begin{figure}[h!]
 \begin{center} 
   \includegraphics[width=10cm]{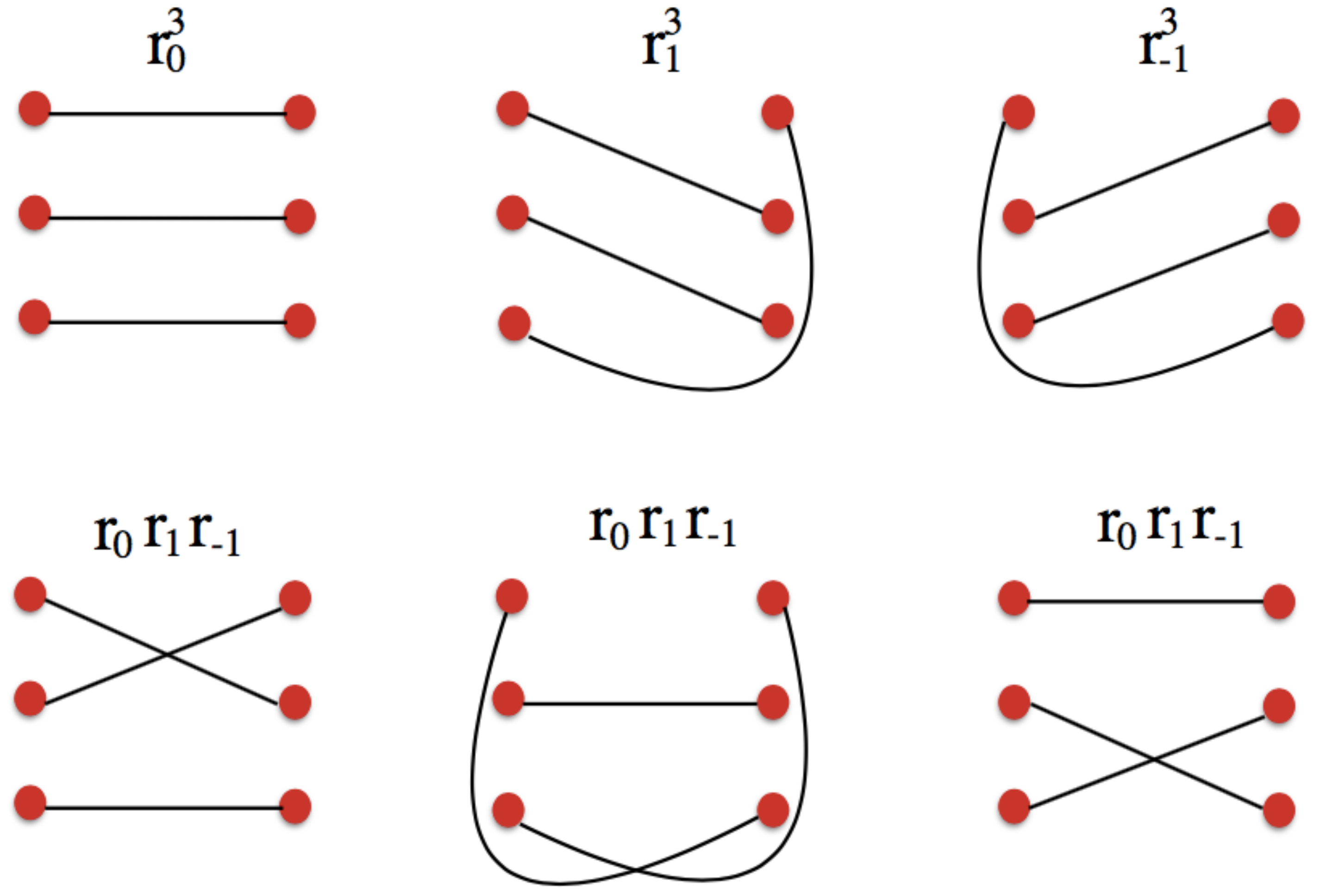} 
  \end{center} 
 \caption{All contributing graphs for $k=1$, $n=3$. In this case the simple addition of individual contributions reproduces the exponential of the function (\ref{negak1n3}).}
 \vspace{1cm}
 \label{casek1} 
 \end{figure}

\begin{figure}[h!]
 \begin{center} 
   \includegraphics[width=10cm]{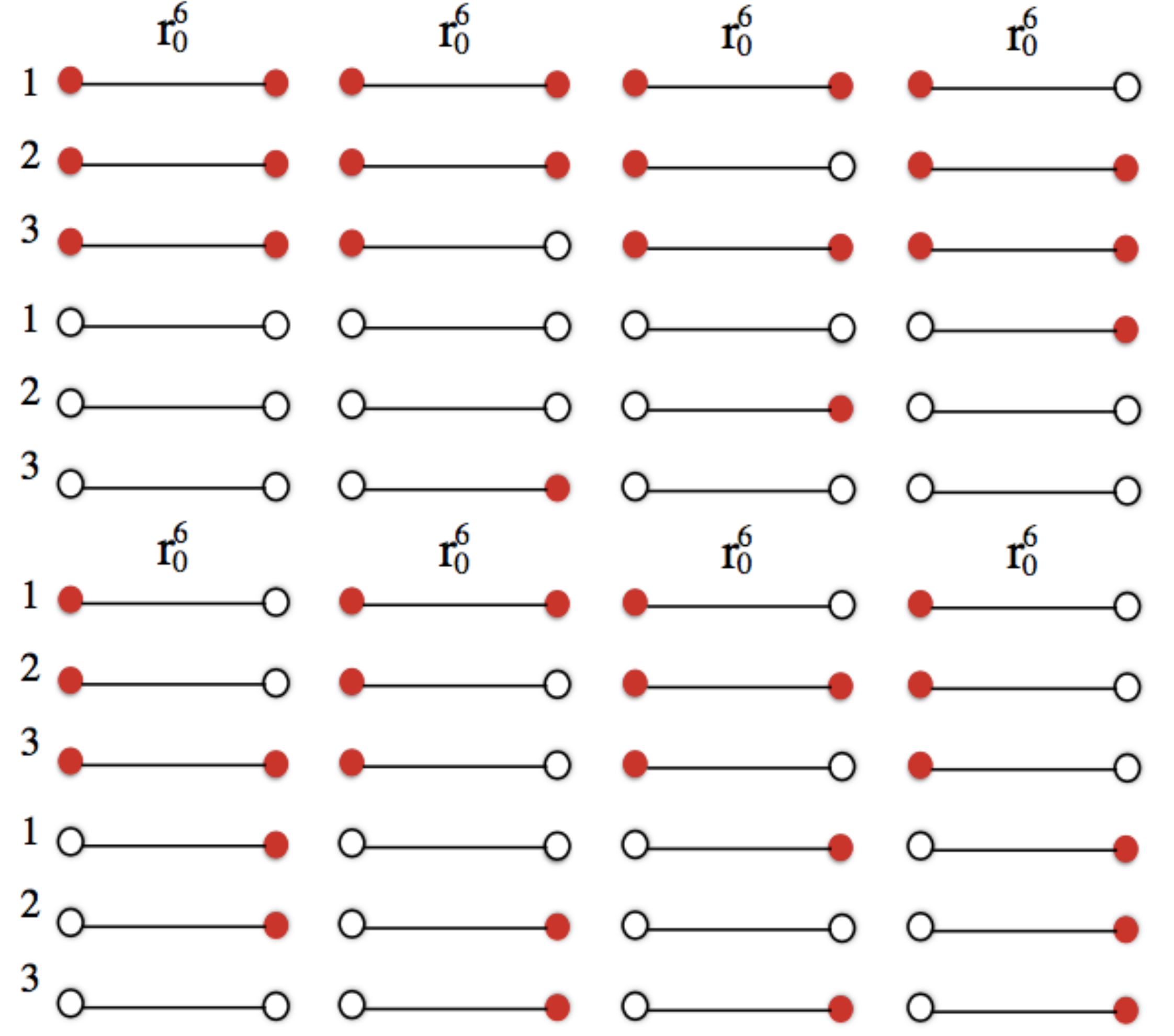} 
  \end{center} 
 \caption{The contribution $r_0^6$ to the negativity for $k=2$, $n=3$. The differently coloured dots represent the two excitations. The numbers on the right hand side label the copies.}
 \vspace{1cm}
 \label{graphsk2} 
 \end{figure}

The contributions discussed above can also be represented graphically. The basic building blocks are the same as for the $k=1$ case, however for $k=2$ every graph may be seen as a ``superposition" of two $k=1$ graphs, and the counting of all possible (non-equivalent) such superpositions that are allowed under the rules set out in Fig.~\ref{negative1} quickly becomes involved. 
Let us consider,  for simplicity, the case $n=3$. The expression for the replica logarithmic negativity is:
\beqa
&& \mathcal{E}_3^{|\Psi_2(r_1,r_0,r_{-1})\ket}=
\label{k2n4}\\
&& \log(r_0^6+ 12 r_1 r_{-1} r_0(r_1^3+r_{-1}^3+r_0^3)+8 (r_1^3 r_{0}^3+ r_{-1}^3 r_0^3+r_1^3 r_{-1}^3)+27 r_1^2 r_{-1}^2 r^2+r_1^6+r_{-1}^6)\,. \nonumber
\eeqa
It is instructive to also report the case $k=1$, $n=3$:
\beq
\mathcal{E}_3^{|\Psi_1(r_1,r_0,r_{-1})\ket}=\log(r_0^3+ 3 r_1 r_{-1} r_0+r_1^3+r_{-1}^3)\,.
\label{negak1n3}
\eeq
Although the expression for $k=1$ is much simpler than for $k=2$, we observe that all contributions to the $k=2$ case can be expressed (apart from the numerical coefficient) as products of two contributions to the $k=1$ case. Thus, whereas for $k=1$ we have a sum of 4 monomials, for $k=2$ we have a sum of 16 monomials. From the point of view of graphs, the case $k=1$, $n=3$ is extremely simple and is represented in Fig.~\ref{casek1}. For $k=2$ we need to normalize each contribution by $\frac{1}{k^n}=\frac{1}{8}$ (representing the fact that each node can now be one of two excitations). For instance, the contribution $r_0^6$ in (\ref{k2n4}) can be seen as the result of adding all graphs in Fig.~\ref{graphsk2} and then dividing the result by 8. 
 More interesting contributions correspond to terms such as $r_1^2 r_{-1}^2 r^2$ which is generated as $(r_1 r_{-1} r_0)^2$, that is as a ``superposition" of any pair of graphs in the second row of Fig.~\ref{casek1}. As in Fig.~\ref{graphsk2} we may have combinations of each of the three graphs with itself, producing $2^3$ possible graphs in each case. However, when  combining two distinct graphs with each other, there are in fact $4^3$ possible combinations for each pairing. Since there are three possible pairings of two distinct graphs and three possible pairings of two identical graphs, this gives $3(2^3+  4^3)=216$ graphs and dividing again by $8$ this gives the coefficient 27 in (\ref{k2n4}).

\section{Recursion for replica negativity from graph partition function, $k=1$ case}\label{sec:recur}

In the case of a single particle excitation ($k=1$), it is possible to write a recursion formula for the graph partition function \eqref{poly}, and by solving it, express the replica logarithmic negativity as the analytic function derived by different means in \cite{the4ofus3}. 
Let us consider a restricted graph partition function 
\beq
\tilde{p}_{1,n}(r_1,r_0,r_{-1}) = \sum_{g\in{\tilde{\mathsf G}}_{1,n}} \prod_{\ell\in\{1,0,-1\}} r_\ell^{N_\ell(g)} \,,
\eeq
summing only for graphs $\tilde{\mathsf G}_{1,n}$, where the vertices $V^{l,r}_{1,n+1}\neq V^{l,r}_{1,1}$ and $V^{l,r}_{1,0}\neq V^{l,r}_{1,n}$ are not identified. This restricts the possible edges of $\mathsf E_1$ and $ \mathsf E_{-1}$. There is no edge in $\mathsf E_1$ attached to $V^l_{1,n}$ and no edge in $\mathsf E_{-1}$ attached to $V^r_{1,1}$.  Moreover for every edge $(V^l_{1,j},V^r_{1,j+1})\in\mathsf E_1$ there is an edge $(V^l_{1,j+1},V^r_{1,j})\in\mathsf E_{-1}$, hence $N_1(g)=N_{-1}(g)$. This restriction give rise to the recursion relation
\beq
\label{eq:rest_part}
\tilde{p}_{1,n}(r_1,r_0,r_{-1})=r_0\tilde{p}_{1,n-1}(r_1,r_0,r_{-1})+r_1 r_{-1} \tilde{p}_{1,n-2}(r_1,r_0,r_{-1})\,.
\eeq 
With the initial conditions $\tilde{p}_{1,0}(r_1,r_0,r_{-1})=1$ and $\tilde{p}_{1,1}(r_1,r_0,r_{-1})=r_0$ the solution of the recursion is 
\beq
\tilde{p}_{1,n}(r_1,r_0,r_{-1})= \frac{\left( r_0 + \sqrt{r_0^2+4r_1 r_{-1}}  \right)^{n+1}-\left( r_0 - \sqrt{r_0^2+4r_1 r_{-1}}  \right)^{n+1}}{2^{n+1}\sqrt{r_0^2+4r_1 r_{-1}}}\,.
\eeq
The original graph partition function can be expressed with the help of the restricted graph partition function and two additional graphs where all the edges are either in $\mathsf E_{1}$ or $\mathsf E_{-1}$
\beq
p_{1,n}(r_1,r_0,r_{-1})=r_1^n+r_{-1}^n+\tilde{p}_{1,n}(r_1,r_0,r_{-1})+r_1r_{-1}\tilde{p}_{1,n-2}(r_1,r_0,r_{-1})\,.
\eeq
Substituting \eqref{eq:rest_part} we arrive to the result 
\beq
p_{1,n}(r_1,r_0,r_{-1})=r_1^n+r_{-1}^n+ \left( \frac{r_0 + \sqrt{r_0^2+4r_1 r_{-1}}}{2}  \right)^{n} + \left( \frac{r_0 - \sqrt{r_0^2+4r_1 r_{-1}}}{2}  \right)^{n}\,,
\eeq
that is the expression for $\exp\big[\mathcal{E}_n^{1}\big]$ derived in \cite{the4ofus3} where each term in the expression above is identified with a non-vanishing eigenvalue of the partially transposed reduced density matrix of the corresponding qubit state. From this formula, the analytic continuation $m\rightarrow 1/2$ for $n=2m$ also follows naturally.


\begin{thebibliography}{99}
\bibitem{bennet}
 C.~H. Bennett, H.~J. Bernstein, S.~Popescu, and B.~Schumacher,
 \newblock Concentrating partial entanglement by local operations,
 \newblock Phys. Rev. {\bf A53}, 2046--2052 (1996).
 
 \bibitem{Eisert}
 K.~Audenaert, J.~Eisert, M.~B. Plenio, and R.~F. Werner,
 \newblock Entanglement properties of the harmonic chain,
 \newblock Phys. Rev. {\bf A66}, 042327 (2002).
 
 \bibitem{ZHSL}
 K.~\ifmmode~\dot{Z}\else \.{Z}\fi{}yczkowski, P.~Horodecki, A.~Sanpera, and
   M.~Lewenstein,
 \newblock Volume of the set of separable states,
 \newblock Phys. Rev. A {\bf 58}, 883--892 (1998).
 
 \bibitem{Eisert2}
 J.~Eisert,
 \newblock Entanglement in quantum information theory (PhD Thesis),
 \newblock quant-ph/0610253  (2006).
 
 \bibitem{ple}
 M.~B. Plenio,
 \newblock Logarithmic Negativity: A Full Entanglement Monotone That is not
   Convex,
 \newblock Phys. Rev. Lett. {\bf 95}, 090503 (2005).
 
 \bibitem{erratumple}
 M.~B. Plenio,
 \newblock Erratum: Logarithmic Negativity: A Full Entanglement Monotone That Is
   not Convex,
 \newblock Phys. Rev. Lett. {\bf 95}, 119902 (2005).
 
 \bibitem{VW}
 G.~Vidal and R.~F. Werner,
 \newblock {Computable measure of entanglement},
 \newblock Phys. Rev. {\bf A65}, 032314 (2002).
 
 \bibitem{CallanW94}
 C.~J. Callan and F.~Wilczek,
 \newblock {On geometric entropy},
 \newblock Phys. Lett. {\bf B333}, 55--61 (1994).
 
 \bibitem{HolzheyLW94}
 C.~Holzhey, F.~Larsen, and F.~Wilczek,
 \newblock Geometric and renormalized entropy in conformal field theory,
 \newblock Nucl. Phys. {\bf B424}, 443--467 (1994).
 
 \bibitem{latorre1}
 G.~Vidal, J.~I. Latorre, E.~Rico, and A.~Kitaev,
 \newblock Entanglement in quantum critical phenomena,
 \newblock Phys. Rev. Lett. {\bf 90}, 227902 (2003).
 
 \bibitem{Latorre2}
 J.~I. Latorre, E.~Rico, and G.~Vidal,
 \newblock Ground state entanglement in quantum spin chains,
 \newblock Quant. Inf. Comput. {\bf 4}, 48--92 (2004).
 
 \bibitem{Jin}
 B.-Q. Jin and V.~Korepin,
 \newblock Quantum spin chain, Toeplitz determinants and Fisher-Hartwig
   conjecture,
 \newblock J. Stat. Phys. {\bf 116}, 79--95 (2004).
 
 \bibitem{Calabrese:2004eu}
 P.~Calabrese and J.~L. Cardy,
 \newblock Entanglement entropy and quantum field theory,
 \newblock J. Stat. Mech. {\bf 0406}, P002 (2004).
 
 \bibitem{entropy} J. L. Cardy, O. A. Castro Alvaredo and B. Doyon,  ``Form factors of branch-point twist fields in quantum integrable models and entanglement entropy", {\em J. Stat. Phys.} 130, 129 (2008).
 
 \bibitem{next}
 B.~Doyon,
 \newblock {Bi-partite entanglement entropy in massive two-dimensional quantum
   field theory},
 \newblock Phys. Rev. Lett. {\bf 102}, 031602 (2009).
 
 \bibitem{fractal}
 O.~A. Castro-Alvaredo and B.~Doyon,
 \newblock {Entanglement entropy of highly degenerate states and fractal
   dimensions},
 \newblock Phys. Rev. Lett. {\bf 108}, 120401 (2012).
 
 \bibitem{disco1}
 P.~Calabrese, J.~Cardy, and E.~Tonni,
 \newblock {Entanglement entropy of two disjoint intervals in conformal field
   theory},
 \newblock J. Stat. Mech. {\bf 2009}(11), P11001 (2009).
 
 \bibitem{BCDLR}
 D.~Bianchini, O.~Castro-Alvaredo, B.~Doyon, E.~Levi, and F.~Ravanini,
 \newblock {Entanglement entropy of non-unitary conformal field theory},
 \newblock J.Phys. {\bf A48}, 04FT01 (2015).
 
 \bibitem{german1}
 F.~C. Alcaraz, M.~I. Berganza, and G.~Sierra,
 \newblock {Entanglement of low-energy excitations in Conformal Field Theory},
 \newblock Phys. Rev. Lett. {\bf 106}, 201601 (2011).
 
 \bibitem{german2}
 M.~I. Berganza, F.~C. Alcaraz, and G.~Sierra,
 \newblock {Entanglement of excited states in critical spin chains},
 \newblock J. Stat. Mech. {\bf 2012}(01), P01016 (2012).
 
 \bibitem{negativity1}
 P.~Calabrese, J.~Cardy, and E.~Tonni,
 \newblock {Entanglement negativity in quantum field theory},
 \newblock Phys. Rev. Lett. {\bf 109}, 130502 (2012).
 
 \bibitem{negativity2}
 P.~Calabrese, J.~Cardy, and E.~Tonni,
 \newblock {Entanglement negativity in extended systems: A field theoretical
   approach},
 \newblock J. Stat. Mech. {\bf 1302}, P02008 (2013).
 
 \bibitem{ourneg}
 O.~Blondeau-Fournier, O.~Castro-Alvaredo, and B.~Doyon,
 \newblock {Universal scaling of the logarithmic negativity in massive quantum
   field theory},
 \newblock J. Phys. {\bf A49}(12), 125401 (2016).
 
 \bibitem{EERev1}
 L.~Amico, R.~Fazio, A.~Osterloh, and V.~Vedral,
 \newblock Entanglement in many-body systems,
 \newblock Rev. Mod. Phys. {\bf 80}, 517--576 (2008).
 
 \bibitem{specialissue}
 P.~Calabrese, J.~Cardy, and B.~Doyon~(ed),
 \newblock {Entanglement entropy in extended quantum systems},
 \newblock J. Phys. {\bf A42}, 500301 (2009).
 
 \bibitem{EERev2}
 J.~Eisert, M.~Cramer, and M.~B. Plenio,
 \newblock Colloquium: Area laws for the entanglement entropy,
 \newblock Rev. Mod. Phys. {\bf 82}, 277--306 (2010).
 
 \bibitem{kniz}
 V.~Knizhnik,
 \newblock {Analytic fields on Riemann surfaces. II},
 \newblock Comm. Math. Phys. {\bf 112}(4), 567--590 (1987).
 
 \bibitem{orbifold}
 L.~Dixon, D.~Friedan, E.~Martinec, and S.~Shenker,
 \newblock The conformal field theory of orbifolds,
 \newblock Nuclear Physics B {\bf 282}, 13--73 (1987).
 
 \bibitem{permutation}
 O.~A. Castro-Alvaredo and B.~Doyon,
 \newblock {Permutation operators, entanglement entropy, and the XXZ spin chain
   in the limit $\Delta \rightarrow -1$},
 \newblock J. Stat. Mech. {\bf 1102}, P02001 (2011).
 
 \bibitem{the4ofus1}
 O.~A. Castro-Alvaredo, C.~De~Fazio, B.~Doyon, and I.~M. Sz\'ecs\'enyi,
 \newblock Entanglement Content of Quasiparticle Excitations,
 \newblock Phys. Rev. Lett. {\bf 121}, 170602 (2018).
 
 \bibitem{the4ofus2}
 O.~A. Castro-Alvaredo, C.~De~Fazio, B.~Doyon, and I.~M. Sz{\'e}cs{\'e}nyi,
 \newblock Entanglement content of quantum particle excitations. Part I. Free
   field theory,
 \newblock JHEP {\bf 2018}(10), 39 (2018).
 
 \bibitem{the4ofus3}
 O.~A. Castro-Alvaredo, C.~De~Fazio, B.~Doyon, and I.~M. Sz{\'e}cs{\'e}nyi,
 \newblock Entanglement Content of Quantum Particle Excitations II. Disconnected
   Regions and Logarithmic Negativity,
 \newblock arXiv:1904.01035 (2019).
 
 \bibitem{PT1}
 B.~Pozsgay and G.~Takacs,
 \newblock {Form-factors in finite volume I: Form-factor bootstrap and truncated
   conformal space},
 \newblock Nucl. Phys. {\bf B788}, 167--208 (2008).
 
 \bibitem{PT2}
 B.~Pozsgay and G.~Takacs,
 \newblock {Form factors in finite volume. II. Disconnected terms and finite
   temperature correlators},
 \newblock Nucl. Phys. {\bf B788}, 209--251 (2008).
 
 
 
 \bibitem{Essler2016}
 F.~H.~L. Essler and M.~Fagotti,
 \newblock Quench dynamics and relaxation in isolated integrable quantum spin
   chains,
 \newblock J. Stat. Mech. {\bf 2016}(6), 064002 (2016).
 \bibitem{DfiniteTff2005} B. Doyon, Finite-temperature form factors in the Majorana theory, J. Stat. Mech. {\bf 2005}, P11006 (2005).
 
 \bibitem{DfiniteTff2007} B. Doyon, Finite-temperature form factors: a review, SIGMA {\bf 3}, 011 (2007).
 
 \bibitem{ChenDoyon2014} Y. Chen and B. Doyon, Form Factors in Equilibrium and Non-Equilibrium Mixed States of the Ising Model, J. Stat. Mech. {\bf 2014}, P09021 (2014)
 
 \end{thebibliography}


\end{document}